\documentclass[prb,a4paper,superscriptaddress,showpacs,amsmath,amssymb,floatfix]{revtex4}

\usepackage{graphicx}
\usepackage{amssymb}
\usepackage{float}
\usepackage{hyperref}

\usepackage{color}

\def\<{{<}}
\def\>{{>}}

\begin{document}
	
\title{Quantum synchronization and entanglement of dissipative qubits \\
  coupled to a resonator}

\author{Alexei D. Chepelianskii}
\affiliation{LPS, Universit\'e Paris-Sud, CNRS, UMR 8502, Orsay F-91405, France}
\author{Dima L. Shepelyansky}
\affiliation{\mbox{Laboratoire de Physique Th\'eorique, 
	Universit\'e de Toulouse, CNRS, UPS, 31062 Toulouse, France}}
	
	\date{August 7, 2022}
	
	\begin{abstract}
	  We study the properties of a driven cavity coupled to several qubits in the framework
          of a dissipative Jaynes-Cummings model. We show that the rotating wave approximation (RWA)
        allows to reduce the description of original driven model to an effective
        Jaynes-Cummings model with strong coupling between photons and qubits.
        Two semi-analytical approaches are developed to describe the steady state of this system.
        We first treat the weak dissipation limit where we derive  perturbative series of rate equations
        that converge to the exact RWA steady-state except near the cavity resonance.
        This approach exactly describes the multi-photon resonances in the system.
        Then in the strong dissipation limit we introduce a semiclassical approximation which allows
        to reproduce the mean spin-projections and cavity state. This approach reproduces the RWA exactly
        in the strong dissipation limit but provides good qualitative trends even in more quantum regimes.
        We then focus on quantum synchronization of qubits through their coupling to the cavity.
        We demonstrate the entangled steady state of a pair of qubits synchronized through their interaction with a driven cavity in 
        presence of dissipation and decoherence.
        Finally we discuss synchronization of a larger number of qubits.
	\end{abstract}

	%

	\maketitle
	
	\section{Introduction} 
	\label{sec1}

        The synchronization of two maritime pendulum clocks,
discovered by  Christian Huygens in 1665 \cite{huygens},
is at the basis of synchronization phenomena abundantly appearing in various systems
ranging from  clocks to fireflies, cardiac pacemakers, lasers and
Josephson junction (JJ) arrays (see historical survey and overview in
\cite{pikovsky,weisenfeld,likharev}).

With the modern development of quantum computation and quantum information
(see e.g. \cite{chuang}) the investigation of quantum synchronization phenomenon
in a quantum world becomes of significant importance.
The phenomenon of quantum synchronization earns
even higher relevance for JJ arrays and superconducting qubits 
where dissipative and quantum effects play an important role
(see e.g. \cite{likharev} and \cite{wendin}).
A strong coupling regime of several superconducting qubits with a microwave resonator
has been realized experimentally \cite{wallraff1,wallraff2,wallraff3,wallraff4,wallraff5}
that requires to investigate also effects of strong interactions
in  quantum dissipative driven systems.

The essential element of synchronization in classical mechanics
is dissipation that leads to a phase synchrony of
coupled autonomous systems with energy supply \cite{pikovsky}.
Thus the analysis of quantum synchronization
requires a use of formalism of dissipative quantum mechanics.
This formalism  is known and based on the Lindblad master equation
for the density matrix $\rho$ of a whole system being developed in
\cite{gorini,lindblad} and reviewed in \cite{weiss}.
It has fundamental grounds and allows to study such complex phenomena as
e.g. quantum strange attractors \cite{graham1987,graham1988}.
However, its numerical simulation requires
integration of time propagation of the whole density matrix with $N \times N$
components that is numerically rather heavy. 

Another approach to dissipative quantum evolution
is based on the method of quantum trajectories \cite{brun1996,brun2002}
which stores only a stochastically evolving state vector of size $N$.
An averaging over several realizations allows to
characterize the probabilities from density matrix with
statistical fluctuations. The application of quantum trajectories
to the problem of quantum synchronization  has been reported 
in the early study \cite{zhirovqsync} where it was shown that
at small dimensionless values of Planck constant $\hbar$
the synchronization  remains robust with respect to quantum
fluctuations. These model studies of quantum synchronization \cite{zhirovqsync}
were shown to have close similarities with Shapiro steps in Josephson junctions
\cite{likharev}. The method of quantum trajectories was also applied
to investigations of quantum synchronization
of one and two qubits coupled to a driven dissipative resonator
\cite{zhirovqubit1,zhirovqubit2}. It was shown that the driving phase
can impose  quantum synchronization
of qubit phases and mutual synchronization and entanglement
of two qubits. In the frame of synchronization
of classical systems this corresponds to the regime
of phase synchronization of oscillators by external driving \cite{pikovdrive}.
At the same time the description with quantum trajectories does not
provide a complete information about
evolution of density matrix. Thus it is important to
study the phenomenon of quantum synchronization of qubits
using the Lindblad master equation for density matrix.
In this work we describe the results obtained
with the Lindlad description.

In last years there is a growing interest to various aspects of
quantum synchronization of various systems
(see e.g. \cite{fazio,bruder1,holland,mavrogordatos,nori,bruder2,bruder3}).
There is even discussion of quantum synchronization for
satellite networks \cite{satellite,satnat2023,satnat2023b}. These developments
initiated our studies of quantum synchronization of
several superconducting qubits coupled to a driven dissipative resonator
performed in the frame of Lindblad master equation.

The performed extensive numerical simulations of tens of thousands of Lindblad
equation components allow to establish various nontrivial regimes
of quantum synchronization and entanglements of one, two, three or four qubits
strongly coupled with a driven resonator in presence of dissipation.
It is shown that in the regimes of weak or strong dissipation
the system behavior can be well described by 
semi-analytical methods  of weak perturbation theory or
semiclassical theory respectively. However,
certain nontrivial dissipative regimes with preserved entanglement
of dissipative qubits remain non-accessible for such a semi-analytical description.

The article is composed as follows: Section II describes the model,
numerical methods and semi-analytical approximations, Section III
presents obtained numerical results and RWA validity tests of RWA,
the results and validity of the weak damping rate equation approach are given in Section IV
and those of strong damping semicalssical regime in Section V,
the regimes beyond the validity of semi-analytical approaches are analyzed in Section VI,
synchronization of several qubits is described in Section VII
and discussion is given in Section VIII.

\section{Model description}

We consider a model of several qubits (spin half systems) interacting with one harmonic cavity
driven by an external monochromatic field. The Hamiltonian of this system is as follows:
\begin{align}
  {\hat H}(t) &= \hbar \omega_0 {\hat a}^+ {\hat a} + \hbar \sum_l \lambda_l {\hat \sigma}_{x,l} ({\hat a}
  + {\hat a}^+) + \sum_l \frac{\hbar \Omega_l}{2} {\hat \sigma}_{z,l} + 2 F ( {\hat a} + {\hat a}^+ ) \cos \omega t  
\label{eq:H0}
\end{align}
where $\omega_0$ is the frequency of the cavity, $\omega$ is the frequency of
the driving field and $F$ is the driving strength. The operators ${\hat a}^+$ and ${\hat a}$ are
the rising/lowering operators for the cavity photons.
The qubits in this model are indexed by $l$ (we consider various numbers of qubits, from 1 to 4),
their contributions to the Hamiltonian are given by the Pauli matrix operators where $\Omega_l$
is the energy splitting (Zeeman splitting for spin systems) and
$\lambda_l$ are their coupling strengths with the cavity.

For $F=0$ and one qubit the Hamiltonian (\ref{eq:H0}) is reduced to the Jaynes-Cummings model \cite{jc}
appearing in many systems of quantum optics  and other physical systems \cite{eberlybook,scully}.
It was experimentally realized with Rydberg atoms in a resonance cavity \cite{walther}.
The unitary behavior of the Jaynes-Cummings model with monochromatic driven cavity
was studied in \cite{jcdriven}. However, here we consider the model with one or several qubits
in presence of driving and dissipation that corresponds to a real situation
of superconducting qubits \cite{wendin}.

In the rotating wave approximation (RWA) the Hamiltonian in Eq.~(\ref{eq:H0}) becomes:
\begin{align}
  {\hat H}(t) &= \hbar \omega_0 {\hat a}^+ {\hat a} + \hbar \sum_l \lambda_l ({\hat a} {\hat \sigma}^+_l
  + {\hat a}^+ {\hat \sigma}^-_l) + \sum_l \frac{\hbar \Omega_l}{2} {\hat \sigma}_{z,l}
  + F ( {\hat a} e^{i \omega t} + {\hat a}^+ e^{-i \omega t}) 
\label{eq:RWA}
\end{align}
where we kept only the resonant terms in both the driving field and spin-cavity interactions.
The Pauli matrices ${\hat \sigma}^+_l$ and ${\hat \sigma}^-_l$ are the rising/lowering spin operators for spin/qubit $l$.

An advantage of the RWA is that the time dependent Hamiltonian Eq.~(\ref{eq:RWA}) can be made stationary by
moving to the rotating frame using the unitary transformation:
${\hat U}(t) = \exp\left(i \omega {\hat a}^+ {\hat a} t + i \omega \sum_l {\hat \sigma}^+_l {\hat \sigma}^-_l t \right)$
which acts on the density matrix ${\hat \rho}$ moving it into the rotating frame through
${\hat \rho_{\cal R}} = {\hat U}(t) {\hat \rho} {\hat U}^+(t)$.
It can be checked that this leads to a stationary rotating frame Hamiltonian \cite{eberlybook,scully}:
\begin{align}
  {\hat H}_{\cal R} &= \hbar (\omega_0-\omega) {\hat a}^+ {\hat a} +
  \hbar \lambda \sum_l ({\hat a} {\hat \sigma}^+_l + {\hat a}^+ {\hat \sigma}^-_l)
  + \sum_l \frac{\hbar (\Omega_l-\omega)}{2} {\hat \sigma}_{l,z} + F ( {\hat a} + {\hat a}^+)
  \label{eq:Hrot}
\end{align}
for simplicity we assumed the same coupling strength between qubit and cavity $\lambda$ for all qubits.

We describe dissipative effects in the framework of the time dependent Lindblad equation \cite{weiss},
with zero temperature dissipative terms for both qubits and cavity dynamics:
\begin{align}
  \partial_t {\hat \rho} &= -\frac{i}{\hbar}[ {\hat H}, {\hat \rho} ] + {\cal L}_d(\rho) 
\label{eq:lindblad}
\end{align}
where ${\cal L}_d$ gives the dissipative part of the Lindblad dynamics:
\begin{align}
  {\cal L}_d({\hat \rho}) &= \gamma \left({\hat a} {\hat \rho} {\hat a}^+ - \frac{1}{2} {\hat a}^+ {\hat a} {\hat \rho}
  - \frac{1}{2} {\hat \rho}  {\hat a}^+  {\hat a} \right)
  + \gamma_{s} \sum_l \left({\hat \sigma}^-_l {\hat \rho} {\hat \sigma}^+_l
  - \frac{1}{2} {\hat \sigma}^+_l {\hat \sigma}^-_l {\hat \rho}
  - \frac{1}{2} {\hat \rho}  {\hat \sigma}^+_l  {\hat \sigma}^-_l \right) 
\label{eq:Ld}
\end{align}
with $\gamma$ being the dissipation rate of the cavity and $\gamma_{s}$ the dissipation rate for qubits.

Under the RWA the Lindblad equation Eq.~(\ref{eq:lindblad}) becomes:
\begin{align}
 \partial_t {\hat \rho}_{\cal R} &= -\frac{i}{\hbar}[ {\hat H}_{\cal R}, {\hat \rho}_{\cal R} ] + {\cal L}_d(\rho_{\cal R})
 \label{eq:RWAlindblad}
\end{align}
since the Hamiltonian ${\hat H}_{\cal R}$ is stationary the steady-state value of
${\hat \rho}_{\cal R}$ can be found by setting the left hand side of Eq.~(\ref{eq:RWAlindblad}) to zero.
The advantage is that finding the steady-state does not require numerical integration
of the equations of motion Eq.~(\ref{eq:lindblad}) and can be found more directly by solving the matrix equation:
\begin{align}
 -\frac{i}{\hbar}[ {\hat H}_{\cal R}, {\hat \rho}_{\cal R} ] + {\cal L}_d(\rho_{\cal R}) = 0
\label{eq:rhoR}
\end{align}

\subsection{Numerical methods}

In general Eq.~(\ref{eq:rhoR}) is a system of linear equations,
with super-operator acting on the unknown density matrix. If the density matrix has size $N \times N$,
the superoperator will be a matrix of size $N^2 \times N^2$ and exploiting its sparse structure of
the Hamiltonian ${\hat H}_{\cal R}$ and of the dissipative Lindblad terms, finding the solution of
Eq.~(\ref{eq:rhoR}) becomes numerically prohibitive. Already building
the sparse super-operator matrix can be numerically demanding in terms of memory. 

Thus we used an alternative approach, exploiting as much as possible the similarity between
the Lindblad equation and Sylvester equations which are matrix equations of the form:
\begin{align}
{\hat A} {\hat X} + {\hat X} {\hat B} = {\hat C}
\end{align}  
where ${\hat A}, {\hat B}, {\hat C}$ are known matrices and ${\hat X}$ is the unknown matrix to be found.
For this class of matrix equations more efficient solution algorithms are available \cite{sylvester},
in our case we used the Bartels-Stewart algorithm \cite{sylvester} which has a $N^3$ computation cost
where $N \times N$ is the size of the density matrix.

The general steady-state Lindblad equation:
\begin{align}
  {\cal L}({\hat \rho}) &=  -i [ {\hat H}, {\hat \rho} ] + \sum_k \left( {\hat L}_k {\hat \rho} {\hat L}^+_k
  - \frac{1}{2} {\hat L}^+_k {\hat L}_k {\hat \rho} - \frac{1}{2} {\hat \rho} {\hat L}^+_k  {\hat L}_k \right) = 0 
\end{align}
is not of Sylvester type (here ${\hat H}$ is the Hamiltonian and ${\hat L}_k$ are dissipative operators).
So a direct application of the Bartels-Stewart algorithm is not possible. We thus split
Lindblad super-operator into a first term which looks like a Sylvester equation and a remainder:
\begin{align}
{\cal L}({\hat \rho}) &= -{\cal L}_{0}({\hat \rho}) +  \epsilon {\cal L}_{1}({\hat \rho}) \\
{\cal L}_{0}({\hat \rho}) &= i  {\hat H} {\hat \rho} + \frac{1}{2} \sum_k  {\hat L}^+_k {\hat L}_k {\hat \rho}
- i {\hat \rho} {\hat H}  + \frac{1}{2} {\hat \rho} \sum_k {\hat L}^+_k  {\hat L}_k \\
{\cal L}_{1}({\hat \rho}) &=  \sum_k {\hat L}_k {\hat \rho} {\hat L}^+_k 
\end{align}
The steady-state density matrix ${\cal L}({\hat \rho}) = 0$ will be a fixed point of: 
\begin{align}
{\hat \rho} = {\cal L}_0^{-1} {\cal L}_1 {\hat \rho}
\end{align}
It is thus possible to iterate from an initial guess ${\hat \rho}_m$ solving a series of Sylvester equations:
\begin{align}
{\cal L}_0 \rho_{m+1} = {\cal L}_1 {\hat \rho}_m
\end{align}
We have found, that usually a few or at most tens of iterations are enough to
find the fixed point with very high accuracy.
The initial guess ${\hat \rho}_0$ for the iteration can be chosen from one of the two approximate methods
presented in the following Sections, we typically used ${\hat \rho}_0$ obtained from summation
of the rate equation perturbation theory.
Due to the availability of a good initial ${\hat \rho}_0$ in our simulations the number of
Sylvester equation iterations required for convergence was small. Algorithms for solving generalized forms
of the Sylvester equation have also been reported \cite{King-wah},
they can maybe offer a faster converging iterative solution in cases where a good initial guess is not available.

It is also possible to find the steady-state by numerical integration of
time dependent Lindblad Eqs.~(\ref{eq:lindblad}),~(\ref{eq:Ld})
or Eq.~(\ref{eq:RWAlindblad}). We used a standard odeint integration library  \cite{odeint} for the integration
of equations of motions using a sparse matrix representation for all operators acting on the density matrix which
is represented as a dense matrix. Main integrator was an adaptive Runge-Kutta-Dorpi fifth order stepping algorithm,
although other integration schemes were also tested with similar results. In the time dependent case it is possible
to check the validity of the RWA approximation since we have access to the full time dependent evolution
of the density matrix including its oscillations around the RWA steady-state, while integration
of the stationary Eq.~(\ref{eq:RWAlindblad}) converges to the solution of Eq.~(\ref{eq:rhoR}).

\subsection{Rate equation perturbation theory}

The above approach for steady-state computation can still be numerically costly if an important number
of Sylvester equation inversions is required. It is thus important to have a good initial guess
for the density matrix to start the iteration. In the limit where the damping terms
in the Lindblad dynamics are weak compared to characteristic frequencies of the Hamiltonian dynamics
it is also natural to use an approximate solution based on a formal expansion of the density matrix
in powers of the dissipation operators. In such an approach the dominant terms of
the density matrix correspond to weighted eigenstates of the Hamiltonian dynamics and
it can be convenient to think of this expansion as a series of rate equations describing
the population of the RWA Hamiltonian eigenstates. We  use this approach
to obtain good initial guesses for density matrix, it  also allows us to investigate
if the steady-state of model Eq.~(\ref{eq:RWAlindblad})  can indeed be described
by a perturbative approach based on Hamiltonian eigenstates. Heuristically,
this is justified only when there is a clear separation of times scales between
the Hamiltonian dynamics and the slower dissipative processes.
Our calculations  allow to justify this point quantitatively. 

We describe this approach starting from the general steady-state Lindblad equation introducing
a formal expansion parameter $\epsilon$ to keep track of the order of the dissipative parameters in the expansion:
\begin{align}
  {\cal L}({\hat \rho}) &=  -i [ {\hat H}, {\hat \rho} ] + \epsilon \sum_k \left( {\hat L}_k {\hat \rho} {\hat L}^+_k
  - \frac{1}{2} {\hat L}^+_k {\hat L}_k {\hat \rho} - \frac{1}{2} {\hat \rho} {\hat L}^+_k  {\hat L}_k \right) \\
&= {\cal L}_{\cal H}({\hat \rho}) +  \epsilon {\cal L}_{\gamma}({\hat \rho}) \\
{\cal L}_{\cal H}({\hat \rho}) &= -i [ {\hat H}, {\hat \rho} ] \\
{\cal L}_{\gamma}({\hat \rho}) &=  \sum_k \left( {\hat L}_k {\hat \rho} {\hat L}^+_k
- \frac{1}{2} {\hat L}^+_k {\hat L}_k {\hat \rho} - \frac{1}{2} {\hat \rho} {\hat L}^+_k  {\hat L}_k \right)
\end{align}
here ${\hat L}_k$ are as previously the relaxation operators and ${\hat H}$
is a steady-state Hamiltonian which in our case is given by RWA Eq.~(\ref{eq:Hrot}).

To find the steady-state we need to solve $ {\cal L}({\hat \rho}) = 0$,
and we proceed making a formal expansion in $\epsilon$: 
\begin{align}
  ({\cal L}_{\cal H} + \epsilon {\cal L}_{\gamma})({\hat \rho}_0
  + \epsilon {\hat \rho}_1 + ... + \epsilon^j  {\hat \rho}_j + ... ) = 0
\label{eq:Lerr}
\end{align}

The equations for the lowest order approximation ${\hat \rho}_0$ are the following:
\begin{align}
  & {\cal L}_{\cal H} {\hat \rho}_0 = 0 \;,\; {\cal P}_{\cal D} {\cal L}_\gamma {\hat \rho}_0 = 0
  \;,\; {\rm Tr}\; {\hat \rho}_0 = 1
\end{align}
where the operator ${\cal P}_{\cal D}$ acts on a density matrix defined in the eigenbasis of ${\hat H}$
by keeping only its diagonal part. The first equation ${\cal L}_{\cal H} {\hat \rho}_0 = 0$ imposes
that ${\hat \rho}_0$ is diagonal in the eigenbasis $|n\rangle$ of the Hamiltonian
${\hat H} = \sum_n \epsilon_n |n\rangle \langle n|$.
Thus  ${\hat \rho}_0$ has the form ${\hat \rho}_0 = \sum_n P_n |n\rangle \langle n|$
where $P_n$ can be interpreted as the probability of eigenstate $|n\rangle$.
The last two equations are thus a rate equation determining $P_n$ and a normalization condition $\sum_n P_n = 1$.

The solution of this equation involves building the matrix:
\begin{align}
{\cal K}_{nm} = \langle m |  {\cal L}_\gamma(|n\rangle \langle n|) | m \rangle .
\end{align}
We then find non zero solutions ${\cal K} |\psi\rangle = 0$,
such solutions always exist since $\langle 1, 1, ..., 1| {\cal K} = 0$.

The recurrence equations for the next orders are:
\begin{align}
&{\cal L}_{\cal H} (1 - {\cal P}_{\cal D}) {\hat \rho}_{j+1} + {\cal L}_\gamma  {\hat \rho}_{j} = 0 
\end{align}
which defines the off-diagonal components of ${\hat \rho}_{j+1}$
(in the eigenbasis of ${\hat H}$) but leaves its diagonal part undefined.

The diagonal part of ${\hat \rho}_{n+1}$ is can be found requiring:
\begin{align}
&{\cal P}_{\cal D} {\cal L}_\gamma  {\hat \rho}_{j+1} = 0 
\end{align}
this can be viewed as higher order corrections to
the rate equations changing the occupation state probabilities $P_n$.

Finally the matrix ${\hat \rho}_{j+1}$ is then made traceless by the transformation 
\begin{align}
{\hat \rho}_{j+1} \rightarrow {\hat \rho}_{j+1} - {\hat \rho}_0 {\rm Tr}\; {\hat \rho}_{j+1} \; ,
\end{align}
this transformation is allowed because the induced error is of higher order:
\begin{align}
{\cal L}( {\hat \rho}_0 {\rm Tr}\; {\hat \rho}_{j+1} ) \sim \epsilon^{j+2}
\end{align}
This can also be viewed as an order by order normalization of the density matrix ${\rm Tr} {\rho} = 1$.

We denote as  ${\bf (R)}$ the numerical result of the summation of these series.
Convergence of this expansion can be checked by evaluating the residual error
in Eq.~(\ref{eq:Lerr}) at each step of the summation checking that the error decreases as the number of terms grow. 
If the series is convergent we continue summation until close to machine precision in the residual error which
is typically reached from the first ten terms, if the series is divergent we stop summation
when the residual error starts to increase.

We find that the direct summation of ${\bf (R)}$ seems to have
a rather small convergence radius requiring  the amplitude  of
the dissipation being small compared to energy level splittings:
\begin{align}
\gamma \ll \min_{i,j} |\epsilon_i - \epsilon_j| 
\end{align}
In practice some of these transitions are almost forbidden by selection rules involving
several spin flips and transitions between several oscillator quanta. We thus try
to improve the radius of convergence of rate equation series by incorporating dissipative terms
into ${\cal L}_{\cal H}$. This is done by observing that ${\cal L}_{\cal H}$
acts on the density matrix $|n\rangle \langle m|$ as:
\begin{align}
{\cal L}_{\cal H}(|n\rangle \langle m|) = i (\epsilon_m - \epsilon_n) |n\rangle \langle m|
\end{align}

We can thus formally include into ${\cal L}_{\cal H}$ all the dissipative terms which
are diagonal in the same basis. This is done by introducing a new (super)operator ${\cal L}_{\gamma{\cal K}}$,
which in the eigenbasis of ${\cal H}$ is defined as follows:
\begin{align}
{\cal L}_{\gamma{\cal K}}({\hat \rho})_{nm} = \left\{
\begin{array}{cc}
  \left[ \langle n| {\cal L}_\gamma(|n\rangle\langle m|) |m\rangle \right]
       {\hat \rho}_{nm}  & \; (m \ne n) \\
  0 &\; m = n
\end{array}
  \right.
\end{align}  

The expansion ${\bf (R)}$ is then performed with the updated operators
${\cal {\tilde L}}_\gamma$ and ${\cal {\tilde L}}_{\cal H}$ given by:
\begin{align}
{\cal {\tilde L}}_\gamma &= {\cal L}_\gamma - {\cal L}_{\gamma{\cal K}} \\
{\cal {\tilde L}}_{\cal H} &= {\cal L}_{\cal H} + {\cal L}_{\gamma{\cal K}}
\label{eq:Lstar}
\end{align}
This approach has some similarity with a Jacobi preconditioning step (see e.g. \cite{jacobi}) which can be used
to improve the radius of convergence of iterative algorithms to solve linear equations.
More physically it corresponds to absorbing some terms of the Lindblad equation
into a non-self adjoint dissipate Hamiltonian.

We will show that this updated approach ${\bf (R)^*}$ improves the radius of convergence
of the perturbation series and for low friction divergence seems to occur only close
to the exact resonance $\omega = \omega_0$ when the RWA Hamiltonian Eq.~(\ref{eq:Hrot})
becomes extremely degenerate. In this case dissipative dynamics can no longer
be described satisfactorily based only on populations of the eigenstates of the Hamiltonian.
Another point of view on the finite convergence radius of the rate equation series is that formally
the series can also be summed for negative values of the dissipation rates which then correspond to gain.
The series on the dissipative and gain sides have the same radius of convergence,
but if the gain is larger than some threshold the oscillator energy will diverge.
This argument provides a maybe more physical explanation for finite convergence radius near resonance.

\subsection{Semiclassical approximation}

While the perturbative rate equation approach works at weak dissipation its convergence fails
in the opposite limit of strong damping. In this regime, an approximate description
of the system steady-state can be obtained from a semiclassical trial density matrix.
The advantage of this approach is that the functional to be minimized can be explicitly
computed and the parameters have a clear interpretation
in terms of cavity coherent states and mean spin orientations.

We start by presenting this functional in the simple model case of a driven cavity,
for which the RWA steady-state is a pure coherent state and thus its minimization gives
the steady-state exactly and then generalize this approach to a cavity coupled to qubits. 

Our approach is to minimize:
\begin{align}
  S = {\rm Tr}\;\left[{\cal L}({\hat \rho})\right]^+{\cal L}({\rho})
  \label{eq:S}
\end{align}
for some trial form of the density matrix.

We first consider the simple model of a driven dissipative cavity in RWA:
\begin{align}
  {\hat H} &= \omega_r {\hat a}^+ {\hat a} + F ( {\hat a}  + {\hat a}^+ ) \\
  {\cal L}({\hat \rho}) &=  \left[ {\cal L}({\hat \rho}) \right]^+ =  -i [ {\hat H}, {\hat \rho} ]
  + \gamma \left( {\hat a} {\hat \rho} {\hat a}^+ - \frac{1}{2} {\hat a}^+ {\hat a} {\hat \rho}
  - \frac{1}{2} {\hat \rho} {\hat a}^+ {\hat a} \right) 
\end{align}
where $\omega_r = \omega_0 - \omega$ is the detuning between
the cavity frequency and the driving field $F$ at frequency $\omega$.

We consider a trial density matrix given by a pure coherent state parameterized
by complex $\alpha = \alpha_x + i \alpha_y$ ($\alpha_{x}$ and $\alpha_y$
are the real and imaginary parts of $\alpha$):
\begin{align}
{\hat \rho}_\alpha = |\alpha\rangle\langle\alpha|
\end{align}
where ${\hat a} |\alpha\rangle = \alpha |\alpha\rangle$.

The functional Eq.~(\ref{eq:S}) can then be evaluated as:
\begin{align}
  S_0(\alpha) = S(|\alpha\rangle\langle\alpha|) =  2 \left[F^2
    + 2 F \alpha_x \omega_r + (\alpha_x^2 + \alpha_y^2) \omega_r^2 \right]
  + 2 F \alpha_y \gamma + \gamma^2 \frac{\alpha_x^2 + \alpha_y^2}{2}
\end{align}
The first bracketed term comes from the Hamiltonian dynamics while the other terms include damping effects.

We can check that the minimum $S_0=0$ is achieved at:
\begin{align}
  \alpha_x  &= -\frac{4 F \omega_r}{\gamma^2 + 4 \omega_r^2} \;,\; \alpha_y
  = -\frac{2 F \gamma}{\gamma^2 + 4 \omega_r^2}
  \label{alphaclassical}
  \end{align}
that is the classical response function of an oscillator at frequency $\omega_0$ excited
at a frequency $\omega$ with a detuning $\omega_r = \omega_0 - \omega$
and a relaxation time $\tau = 2 \gamma^{-1}$. Since $S_0 = 0$ this solution is exact.

We now generalize this approach to the case of a cavity coupled to one and then more qubits.
In the case of one-qubit with RWA Hamiltonian Eq.~(\ref{eq:Hrot})
we generalize the trial form of the density matrix to a product state:
\begin{align}
  {\hat \rho} = \frac{1}{2} |\alpha\rangle \langle \alpha|\left(1 + b_{x} {\hat \sigma}_{x}
  + b_{y} {\hat \sigma}_{y} + b_z {\hat \sigma}_{z} \right)
\end{align}
which is now parameterized by three real numbers $b_{x,y,z}$ given
the Bloch sphere coordinates of the spin density matrix, in addition
to the complex number $\alpha$ parametrizing the cavity coherent state.

Evaluating Eq.~(\ref{eq:S}) we find:
\begin{align}
  S =  \frac{1+b_x^2+b_y^2+b_z^2}{2} S_0(\alpha_x, \alpha_y)
  + \frac{\Omega_r^2 (b_x^2+b_y^2)}{2} + S_\lambda + S_{\gamma_s}
\label{eq:semiS}
\end{align}
where the first term corresponds to the single cavity described previously.
In the second term the frequency $\Omega_r = \Omega - \omega$ gives the detuning between
the qubit level spacing $\Omega$ and the driving frequency. Here we omit the qubit index
$\Omega = \Omega_l$ and $\lambda$ is the qubit cavity interaction strength.
This term just describes the qubit eigenstates without qubit-cavity coupling corresponding to $b_{x,y} = 0$. 

The third term $S_\lambda$ describes the Hamiltonian part
of the qubit-cavity coupling and all terms are proportional to $\lambda$.
\begin{align}
  S_{\lambda} &= 2 \lambda \left[ F b_x + (\alpha_x b_x - \alpha_y b_y) (\omega_r - b_z \Omega_r) \right]
  + \gamma \lambda (\alpha_y b_x + \alpha_x b_y)
\\ \nonumber
& +\frac{\lambda^2}{2} \left[4 b_z^2 \left(\alpha_x^2+\alpha_y^2\right)
  +\left(4 \alpha_x^2+1\right) b_y^2+8 \alpha_x \alpha_y b_x b_y+\left(4 \alpha_y^2+1\right) b_x^2+(b_z+1)^2\right]
\end{align}

The final term $S_{\gamma_s}$ contains all terms which are induced by the qubit damping:
\begin{align}
  S_{\gamma_s} =  \gamma_s^2 \frac{b_x^2 + b_y^2 + 4 (1 + b_z)^2}{8}
  - \gamma_s \lambda (\alpha_y b_x + \alpha_x b_y) (2 + b_z) 
\end{align}
where the first term is minimized by the qubit ground state $b_x,b_y = 0$ and $b_z = -1$
while the second term mixes qubit-cavity coupling and qubit relaxation.

The cavity-qubit steady-state is then obtained minimizing $S$ as function
of five parameters $\alpha_{x,y}, b_{x,y,z}$. For the case of two-qubits
we generalize the trial form of the density matrix to:
\begin{align}
  {\hat \rho} = \frac{1}{4} |\alpha\rangle \langle \alpha|\left(1 + b_{1x} {\hat \sigma}_{1x}
  + b_{1y} {\hat \sigma}_{1y} + b_z {\hat \sigma}_{1z} \right)
  \left(1 + b_{2x} {\hat \sigma}_{2x} + b_{2y} {\hat \sigma}_{2y} + b_z {\hat \sigma}_{2z} \right)
  \end{align}
the expression for $S$ for this trial function is given in the Appendix.

Our trial form for the density matrix does not include any entanglement between qubits
and cavity since the trial density matrix is taken as a tensor product. The cavity steady-state
is given by a coherent state and the qubits by their Bloch sphere components. We thus call
this the semiclassical approximation to the cavity-qubit steady-state. While minimization of
the semiclassical function cannot no longer be done analytically for the general case,
this minimization is very light computationally compared to full quantum calculations.
It is thus important to identify the regimes of the Lindblad equation where
the above semi-analytical approaches are able to correctly reproduce its
full quantum dissipative solution.

\section{Numerical and semi-analytical results}

In the previous Section, we described how the time dependent dissipative quantum problem
of a driven cavity coupled to qubits can be transformed into a stationary problem
by moving to the rotating frame. Finding the steady-state density matrix of the system is
then reduced to find the kernel of the RWA Lindblad operator ${\cal L}({\hat \rho})$.
We then presented several approaches to finding this steady-state.
An exact numerical approach based on an iteration of Sylvester equation solutions,
a weak damping perturbation theory series which can be interpreted as population dynamics
of the quantum eigenstates of the system and a semicalssical approximation which we expect
to hold at stronger damping here quantum coherence are quickly destroyed.

Here, we first present some numerical results  confirming the convergence
of the system to its RWA steady-state and then investigate the different regimes
of this model where our two semi-analytical approximations can apply.
We find that for most parameters either one of these approximations can be used
but we also identify a regime where they
are not able to describe the entangled dissipative state of the system.

\subsection{RWA validity analysis}

To probe the validity of the RWA steady-state we consider an example of Lindblad dynamics
for a single non dissipative qubit coupled to a dissipative cavity.
This model has been analyzed previously in \cite{zhirovqubit1} using the method of quantum trajectories.
This method showed a bistability of the quantum trajectory wave-function corresponding
to two qubit orientations even for relatively large detuning between the qubit and cavity eigenfrequencies
compared to their coupling strength $\lambda$.
This nontrivial regime seems to be  a good test-case to investigate RWA validity.

\begin{figure}[!htb]
\centering 
\includegraphics[width=0.8\columnwidth]{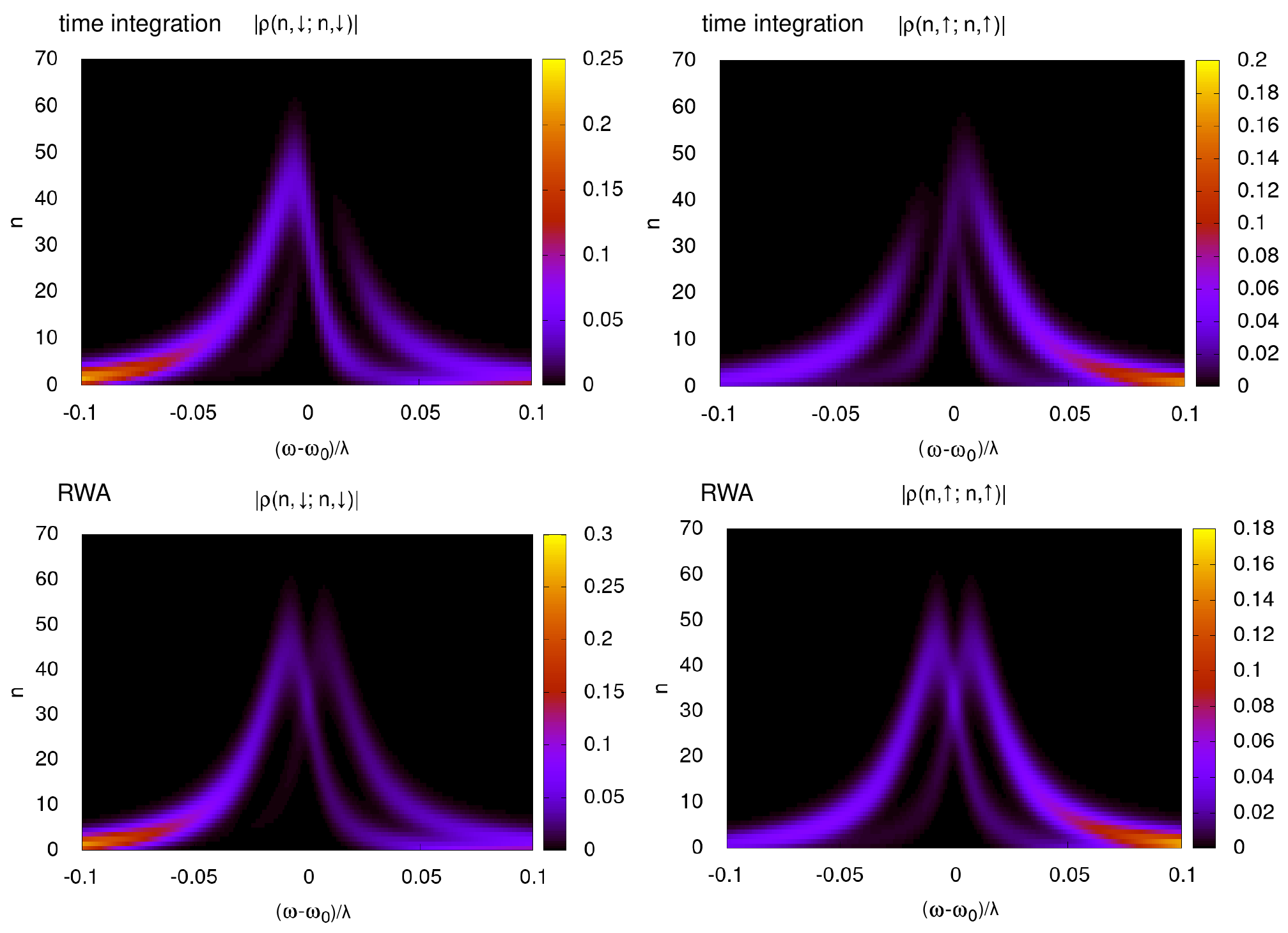} \\
\caption{Distribution of the oscillator occupation number $n$ as function of the detuning from cavity resonance
  $(\omega-\omega_0)/\lambda$ and $z$-axis spin projection. The color scale shows the amplitude $|\rho(n,s;n,s)|$
  where $s = \uparrow, \downarrow$ is the spin up/down state. We remind that $\lambda$ is the strength of the spin cavity interaction,
  we use it as the energy scale to facilate comparison between direct time integration and RWA which depends only
  on the difference $\Delta = \Omega-\omega_0$ between the Zeeman splitting $\Omega$ and the cavity frequency $\omega_0$, which is set to $\Delta = 2\lambda$ in this figure. Here, the driving strength is set to $F = \lambda$, cavity dissipation is $\gamma = 0.3 \lambda$ and qubit is dissipationless $\gamma_s = 0$.
  Top row figures are obtained by direct time integration for $\omega_0/\lambda = 10$,
  while bottom row are results from RWA. For the Lindblad time integration, numerical data shows the density matrix
  after $\tau_p = 2\times 10^4$ excitation periods starting from the system ground-state for $F=0$ at $t=0$
  (we use the same $\tau_p$ for results presentation of other cases of time dependent Lindblad equation).
  The oscillator phase space in the simulations was truncated to the first 100 oscillator levels (usually used for other cases).
}
\label{fig:RWA1}
\label{fig1}
\end{figure}

The bistability discussed in \cite{zhirovqubit1} manifests as two branches of the mean cavity occupation
$\langle a^+ a\rangle$ for the two possible qubit eigenstates. This can be seen as two distinct peaks
in the  probability distribution of the cavity occupation number for spin down
$P_-(n) = \langle n, -| {\hat \rho} | n, -\rangle$ and spin up respectively
$P_+(n) = \langle n, +| {\hat \rho} | n, +\rangle$, with $n$ the cavity eigenstate index.
This bistability is reproduced by integration of time dependent Lindblad evolution
in Fig.~\ref{fig:RWA1} in the top two panels showing $P_-(n)$ and $P_+(n)$ as function
of the detuning $\omega_r = \omega_0 - \omega$ between the cavity and the driving field.
Other parameters are fixed to: $\Delta = 2 \lambda$, where we introduce
$\Delta = \Omega - \omega_0$ as the detuning between qubit and the cavity frequencies,
$F = \lambda$ and $\gamma = 0.3 \lambda$. The cavity frequency is fixed to $\omega_0/\lambda = 10$.
For convenience here, we consider units where $\hbar = 1$ and
use the cavity qubit coupling strength $\lambda$ as the energy scale in dimensionless quantities.
This choice of units is convenient to analyze RWA validity because only relative energies
to the excitation frequency appear in Eq.~(\ref{eq:Hrot}) and thus energy differences rather
than the absolute energy become physically relevant.  Bottom panels in Fig.~\ref{fig:RWA1}
show the same quantities for the RWA steady-state with good agreement between both methods. 

While RWA qualitatively reproduces the bistability behavior, some deviations are also visible.
To be more quantitative on the agreement between RWA and exact dynamics we consider how
the mean qubit/cavity quantities dependence on $\omega_r = \omega_0 - \omega$ are changed with increasing RWA parameter
$\omega_0/\lambda$ which is expected to determine the RWA validity. 

\begin{figure}[!htb]
\centering 
\includegraphics[width=0.8\columnwidth]{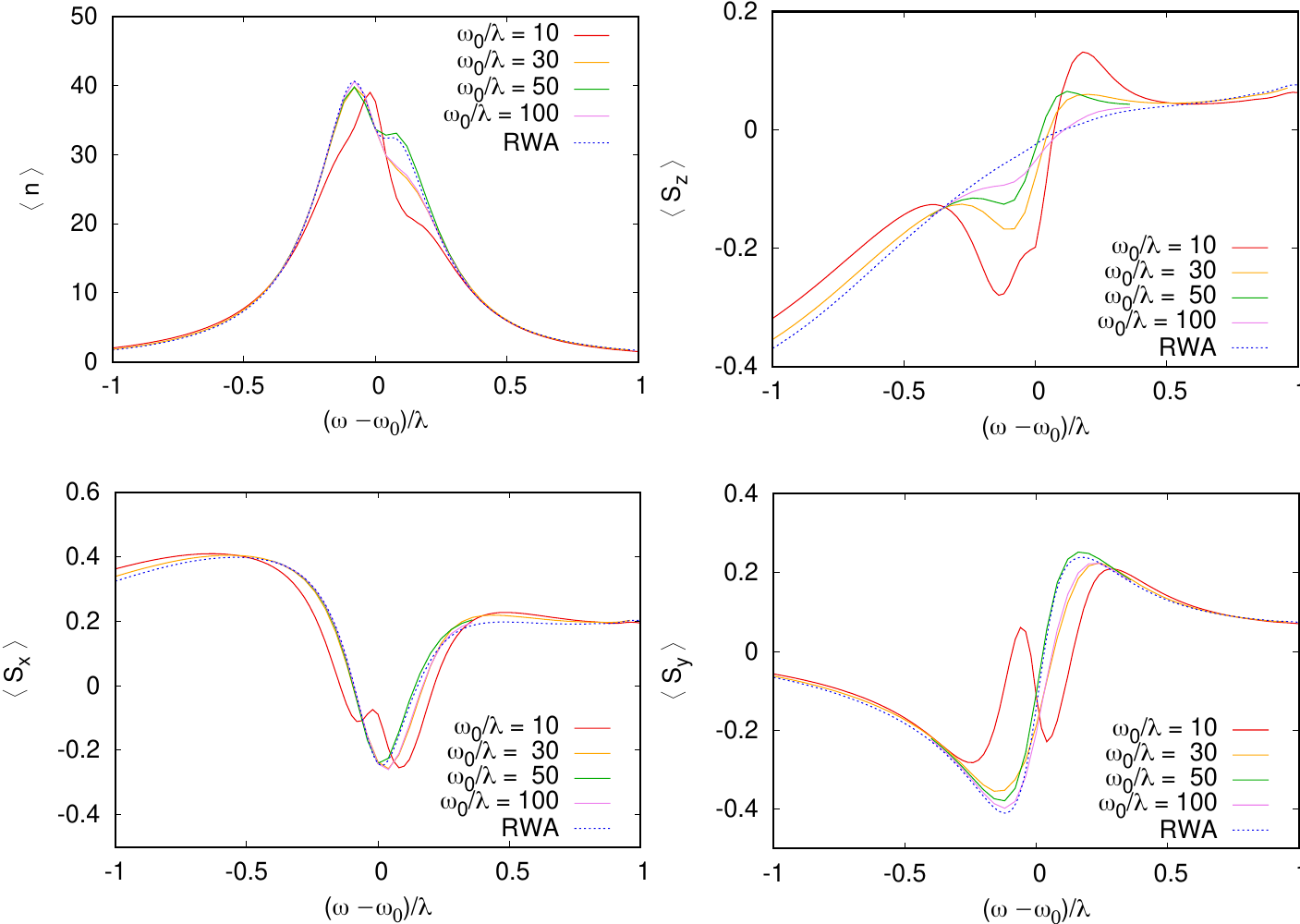} \\
\caption{Mean spin projections $\langle S_{x,y,z} \rangle$ and oscillator quantum number $\langle n \rangle = \langle a^+ a\rangle$
  as function of the detuning  $(\omega-\omega_0)/\lambda$ for $\Omega - \omega_0 = 2 \lambda$, $F = \lambda$, $\gamma = 0.03 \lambda$
  (same values as in Fig.~\ref{fig1}). Different traces correspond to quantum dynamics for increasing RWA parameter
  $\omega_0/\lambda = 10, 30, 50, 100$. The agreement with RWA improves as $\omega_0/\lambda$ increases but worsens
  a bit for the largest value. This behavior is explained as a transient effect in Appendix Fig.~\ref{figA1},
  where we show that relaxation to steady-state was not completed even after $\tau_p=2\times10^4$ microwave periods.
  Indeed since the qubit is not dissipative in this simulation ($\gamma_s = 0$)
  relaxation time scale can be much longer than $\gamma^{-1}$. 
}
\label{fig:RWA2}
\label{fig2}
\end{figure}

Results are shown in Fig.~\ref{fig:RWA2} for similar parameters as Fig.~\ref{fig:RWA1}.
While some deviations between integration of time dependent Eq.~(\ref{eq:H0}) and RWA are visible
for $\omega / \lambda = 10$, the deviation rapidly drops as $\omega/\lambda$ is increased.
For $\langle{\hat S}_x\rangle$,  $\langle{\hat S}_z\rangle$ and  $\langle{\hat a}^+ {\hat a}\rangle$
the convergence is rapid with identical results for $\omega/ \Delta = 10$ already,
while the qubit projection ${\hat S}_y$ converges more slowly. In the numerical simulations
we use  $100$ cavity levels which was found sufficient being twice larger than the maximum excitation
number in Fig.~\ref{fig:RWA1}. Slow relaxation of one of the spin components
in this model is shown in the Appendix Fig.~\ref{figA1}. RWA directly computes the steady-state of the system and
thus does not require to integrate many oscillation periods before reaching steady-state.
The deviations between the RWA steady-state  and fully relaxed time integration
are small in limit $\omega/\lambda \gg 1$  and thus
the RWA can be a more reliable way to explore steady-state properties of this regime.
Part of the deviations between RWA and full time dependent dynamics
can be attributed to the slow relaxation of the spin degrees of freedom in this model, and
thus integration over long times are needed to reach the system steady-state.
Thus, overall  we find that the RWA reproduces the bistability effect quite accurately.
In the framework of quantum trajectories description,
this bistability becomes visible as long time jumps between quantum states,
these jumps  on average should reproduce the populations of each bistable state.

\section{Weak damping rate equation approach}

\begin{figure}[!htb]
\centering 
\begin{tabular}{c}
\includegraphics[width=0.6\columnwidth]{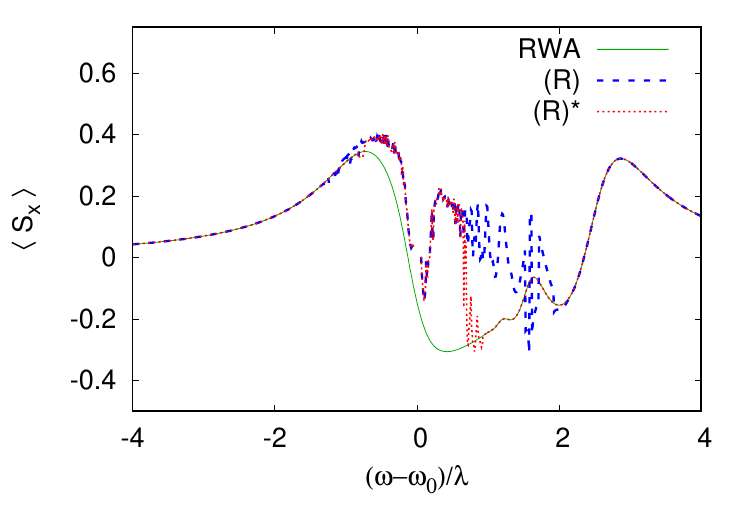}
\end{tabular}
\caption{Comparison of RWA simulation with the summation of rate equation series
  for $F = \lambda$, $\Omega - \omega_0 = 2 \lambda$ and $\gamma = \gamma_S = 0.3 \lambda$.
  The trace (R) corresponds to summation of the series from the direct rate expansion Eq.~(\ref{eq:Lerr})
  while (R)$^*$ exhibiting a larger radius of convergence corresponds to Eq.~(\ref{eq:Lstar}).
  The series (R) qualitatively reproduce the position of the multiphoton resonances but
  with excessive amplitude and fails to converge. The series (R)$^*$ reproduce multiphoton resonace accurately
  but still fail to converge close to the cavity resonance
  $(\omega - \omega_0)/\lambda \sim 1$ (further studies are needed to know
  if divergence occurs on energy scale $\lambda$ or $\gamma$ around the resonance). 
}
\label{fig:ratevsrwa}
\label{fig3}
\end{figure}

We then present a typical example of comparison between the RWA steady-state and the summation
of the rate equation series. We show the comparison for the case of a single qubit interacting
with a cavity in the case where both spin and cavity are dissipative.  Fig.~\ref{fig:ratevsrwa}
shows the mean qubit spin-projection $\langle S_x \rangle$ in the steady-state as function
of the detuning $\omega_r$. Similar results are obtained for other spin components $\langle S_{y,z} \rangle$.
The RWA steady-state found using a Sylvester equation iterations shows series of oscillations
when $\omega_r$ is in the interval $[0, \Delta]$. These oscillations correspond to multiphoton resonances
in the rotating frame $k \omega_r = \Delta$ (where $k$ is a positive integer). In the laboratory
frame these resonances correspond to transitions $(k+1) \omega = k \omega_0 + \Omega$ where $k+1$ photons
are absorbed to excite $k$ cavity levels and the qubit at energy $\Omega$.
The direct rate equation series ${\bf (R)}$ converge only far from the resonance and
fail to reproduce the multiphoton transitions but the corrected series ${\bf (R)^*}$ reproduce
this behavior correctly and fail to converge only in a close vicinity
to the cavity-driving field resonance when $\omega_r \sim \gamma$. 

\begin{figure}[!htb]
\centering 
\begin{tabular}{cc}
\includegraphics[width=0.6\columnwidth]{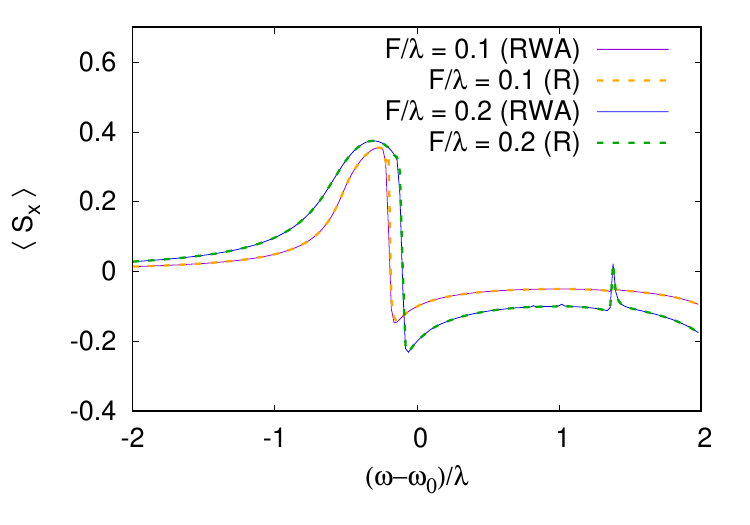}
\end{tabular}
\caption{Comparison between RWA and rate equation series for weak damping
  $\gamma = \gamma_S = 0.005 \lambda$. As expected in this regime,
  the rate equation series (R), from Eq.~(\ref{eq:Lerr}), converges. 
  Excitation was reduced to avoid overheating at resonance with $F = 0.1 \lambda$ and $0.2 \lambda$.
  As in the previous figures $\Omega - \omega_0 = 2 \lambda$.
}
\label{fig:ratevsrwa2}
\label{fig4}
\end{figure}

The comparison in Fig.~\ref{fig:ratevsrwa} is performed at relatively high frictions from
the point of view of the rate equation perturbation theory and we see that the multi-photon resonances make
the direct rate equation series ${\bf (R)}$ unstable in a wide range of $\omega_r$ highlighting
the requirement $\gamma \ll \min_{i,j} |\epsilon_i - \epsilon_j|$. For weaker damping
the rate equation series converge in a much wider range of detuning $\omega_r$ as shown
on Fig.\ref{fig:ratevsrwa2}. In this case even ${\bf (R)}$ converges almost everywhere except at
very narrow resonances where some deviation from the exact steady-state is still visible.

\section{Strong damping semiclassical regime}

\begin{figure}[!htb]
\centering 
\begin{tabular}{cc}
\includegraphics[width=0.5\columnwidth]{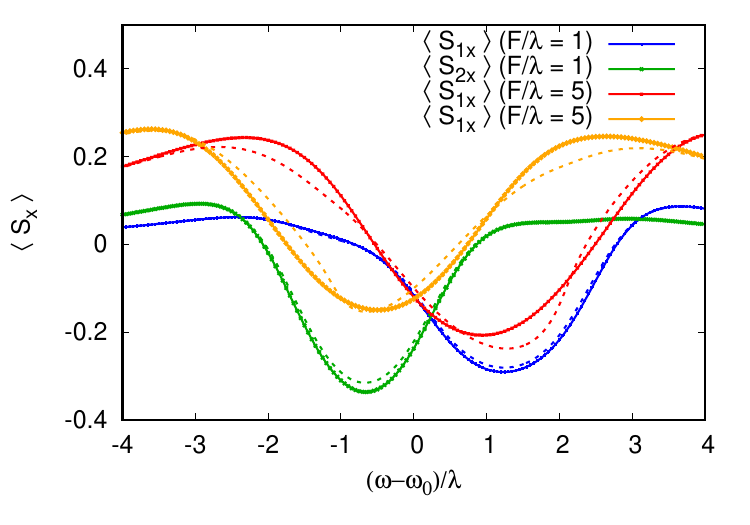} & \includegraphics[width=0.5\columnwidth]{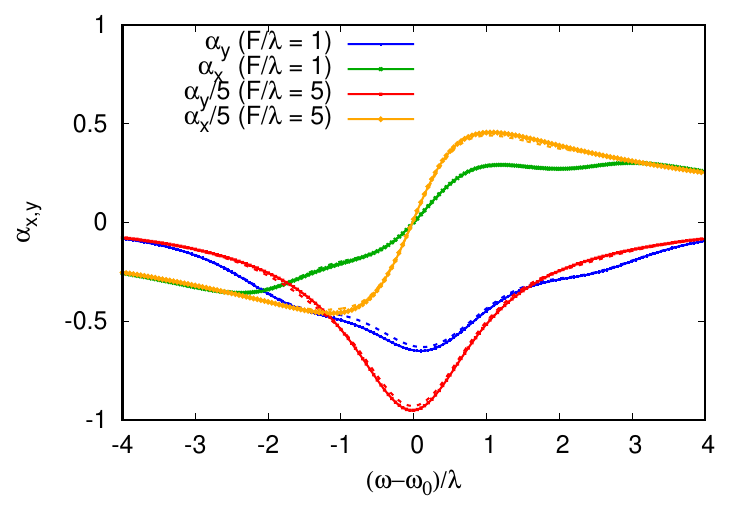}   
\end{tabular}
\caption{ Spin projections of the two spins as function of the detuning $(\omega-\omega_0)/\lambda$
  for excitation strength $F= \lambda$ and $F = 5 \lambda$. Dashed lines show the spin projection
  predicted by the semiclassical functional Eq.~(\ref{eq:semiS}), doted lines show the RWA steady state.
  The qubit-cavity detunings are: $\Delta_1 = \Omega_1 - \omega_0 = 2 \lambda$,
  $\Delta_2 = \Omega_2 - \omega_0 = -\lambda$.
  The dissipation is fixed to $\gamma = \gamma_{s} = 2 \lambda$, the relatively large value of
  the dissipation rates ensures good agreement with the semiclassical predictions.
}
\label{fig:semivsrwa}
\label{fig5}
\end{figure}

We find that in the opposite limit of strong damping the semiclassical approximation is quite successful.
To illustrate this we consider a cavity coupled to two qubits with all dissipation terms
in Eqs.(\ref{eq:lindblad},\ref{eq:Ld}) both for cavity and qubits with equal dissipations
$\gamma = \gamma_s$. In this case the dissipation
is chosen to be of the same order of magnitude as the detuning between qubits and cavity.
This is thus a regime where we expect the semicalssical theory
to give a very good description of the steady-state.
This is confirmed by numerical simulations presented in Fig.~\ref{fig:semivsrwa}
which shows that both  the mean spin orientation and the mean cavity susceptibility variables
$\alpha_x = {\rm Re} \langle {\hat a} \rangle$ and $\alpha_y = {\rm Im} \langle {\hat a} \rangle$
are reproduced almost perfectly by the minimization of the semiclassical functional $S$.
We see that in addition to the main resonance at $\omega_r = 0$ the cavity susceptibility
shows also features at resonance with the qubits $\omega = \Omega_{1,2}$ in agreement with exact RWA solution.
For these strong values of  damping the rate equations series are
no longer converging for the range of detunings explored here.

\begin{figure}[!htb]
\centering 
\includegraphics[width=0.5\columnwidth]{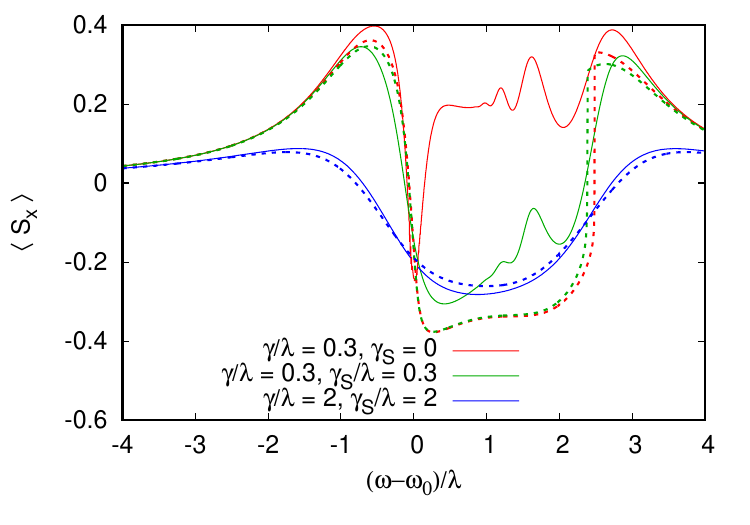}
\caption{This figure compares the spin polarization  $\langle S_x \rangle$ for one qubit coupled
  to a cavity for increasing dissipation rates. The polarization is shown as function of
  $(\omega - \omega_0)/\lambda$ for $F = \lambda$, $\Omega - \omega_0 = 2 \lambda$. Smooth thin curves show
  the RWA steady-state while dashed curves show the semi-classical theory. For the lowest dissipation
  $\gamma_s = 0, \gamma = 0.3 \lambda$ the semiclassical theory predicts a reversal of  $\langle S_x \rangle$
  in the range $(\omega - \omega_0)/\lambda \in (0, 2)$ which is not present in RWA which gives
  the exact quantum result but agreement is good outside this range. Adding some dissipation to
  the spin $\gamma_s = \gamma = 0.3 \lambda$ is enough to observe the polarization reversal in RWA also.
  Agreement becomes very good for $\gamma_s = \gamma = 2 \lambda$.
  As expected at a higher friction the system behaves in a more semiclassical way. 
}
\label{fig:semivsrwa2}
\label{fig6}
\end{figure}

At weaker frictions the semiclassical results become less accurate but can reproduce
still qualitative trends of the qubit response to cavity excitation. To illustrate
this we show the semiclassical results for the case from Fig.~\ref{fig6} for one qubit.
We see that although this method fails to capture the multiphoton resonance
the broad shape of the response is reproduced correctly. 

\section{Beyond semiclassical analysis and rate equations}

We found that the rate equation and semiclassical approaches capture complementary regimes of
the dissipative quantum dynamics with respectively low and high damping regimes.
It is thus interesting if we can find a regime not accessible to these approaches
where exact solution of dissipative quantum dynamics is required. This regime would have to be at a weak damping
near resonance drive since this is the only regime which is outside the range of validity of both approaches.
However, a strong cavity excitation at a resonance can lead to
an effectively semiclassical overheated regime, without special quantum coherence properties.

We thus consider the following setting where two qubits are coupled to the cavity
with opposite detunings from the cavity frequency $\Delta_1 = -\Delta_2$ (where $\Delta_1 = \Omega_1 - \omega_0$
and $\Delta_2 -(\Omega_2 - \omega_0)$ are the qubit-cavity detunings). The cavity excitation  at
frequency $\omega = \omega_0$ does not break the symmetry between the qubits in RWA,
and this seems a good regime to induce entanglements between qubits even in presence of damping.
In the following we analyze this model in detail with RWA approach (\ref{eq:RWAlindblad})
and exact time dependent Lindblad dynamics (\ref{eq:lindblad}),~(\ref{eq:Ld}).
We also consider the validity of our two approximate approaches in this regime.

\begin{figure}[!htb]
\centering 
\includegraphics[width=0.6\columnwidth]{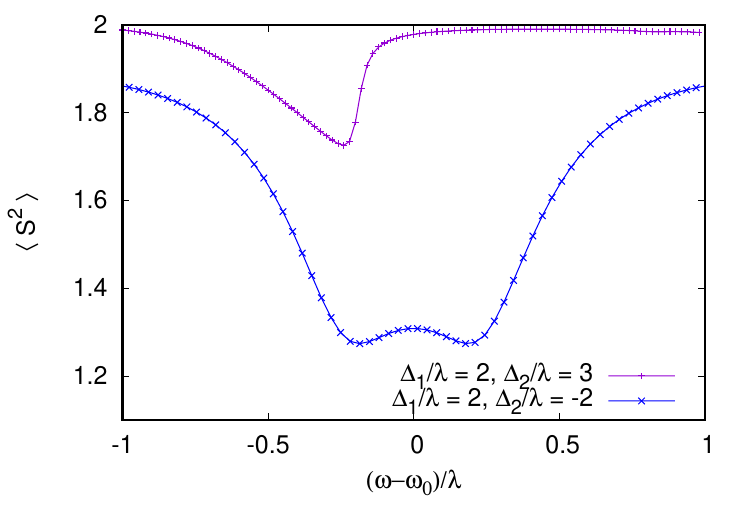} 
\caption{ RWA calculation of the total spin of the qubit pair as function of $(\omega-\omega_0)/\lambda$
  for two values of the qubit-cavity detuning $\Delta_{1,2} = \Omega_{1,2} - \omega_0$. The top curve corresponds
  to $\Delta_1/\lambda = 2$ and $\Delta_2/\lambda = 3$ with only a weak deviation from the equilibrium
  total spin triplet $\langle S^2 \rangle = S(S+1) = 2$. When the two Zeeman splittings are antisymmetric
  with respect to the cavity $\Delta_1 = -\Delta_2 = 2\lambda$
  a significantly stronger reduction of $\langle S^2 \rangle$ is observed.
  Dissipative rates are set to $\gamma = \gamma_S = 0.2 \lambda$ and excitation is $F = 1.5 \lambda$.
  As previously 100 oscillator levels are used in the simulation.}
\label{fig:Singlet1}
\label{fig7}
\end{figure}

To monitor the entangled state of the two detuned two qubits we consider their mean total spin operator ${\hat S}^2$
whose expectation value $S(S+1) = 2$ for the triplet $S=1$ state of the two qubits and zero for the singlet configuration.
The dependence of $\langle {\hat S}^2 \rangle$ on $\omega-\omega_0$ is shown on Fig.~\ref{fig:Singlet1}
for two cases of qubit detuning from the cavity, comparing the antisymmetric detuning introduced
above and and more generic values with $(\Omega_1 - \omega_0) / \lambda = 2, (\Omega_2 - \omega_0) / \lambda = 3$.
For antisymmetric detuning a strong reduction of $\langle {\hat S}^2 \rangle$ is observed
when the cavity is excited at resonance, this corresponds to a dynamics of the two qubits
where they rotate together, synchronizing in opposite directions and canceling
(at least partially) their total spin. This reduction is not observed for the generic detuning values
for which  $\langle {\hat S}^2 \rangle$ stays close to 2, which is its equilibrium value. 

\begin{figure}[!htb]
\centering 
  \includegraphics[width=0.6\columnwidth]{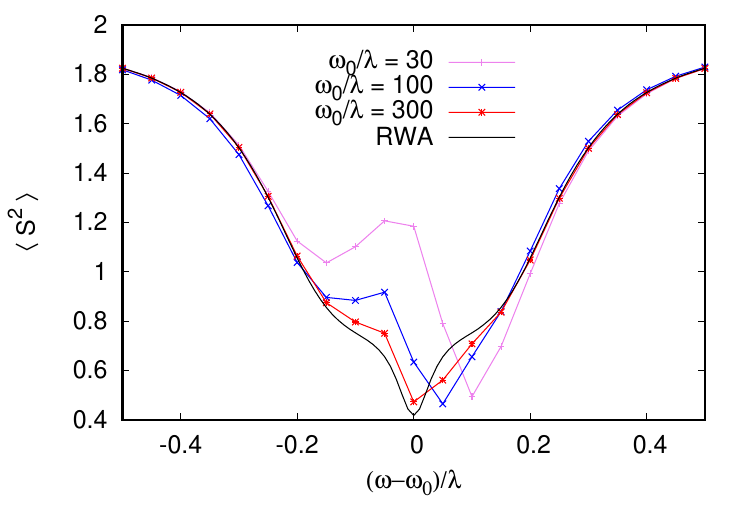} 
  \caption{ When dissipative rates are reduced compared to Fig.~\ref{fig7}, the reduction of $\langle S^2 \rangle$
    for antisymmetric qubit-cavity detunings $\Delta_1 = -\Delta_2 = \lambda$ becomes stronger and the singlet state
    of the qubit pair becomes the most probable  state (75\% singlet probability at the minimum of $\langle S^2 \rangle$).
    Here, the driving strength is set to $F/\lambda = 0.25$, the dissipative rates are $\gamma = 0.3 \lambda$
    with a weak qubit dissipation $\gamma_S = 10^{-3} \gamma$ (for these parameters, at resonance,
    $ \langle n \rangle \simeq 4 F^2/\gamma^2 \simeq 3$). To confirm that the singlet formation
    is not an artifact of the RWA (black curve),
    we performed direct integration of the time dependent Lindblad dynamics up to total simulation time
    $3 \gamma_s^{-1}$ for increasing RWA parameter $\omega_0/\lambda$ (color curves with symbols).
    The singlet formation is robust to non RWA effects
    with the minimum $\langle S^2 \rangle$ remaining unchanged as $\omega_0/\lambda$ is varied by an order
    of magnitude. Only weak Non RWA effects are visible as a small shift of the minimum from $\omega = \omega_0$
    and an asymmetric $\langle S^2 \rangle$ dependence, since non RWA effects break
    the symmetry between two anti-symmetrically detuned qubits.
}
\label{SingletRWAvsNonRWA}
\label{fig8}
\end{figure}

Fig.~\ref{fig:Singlet1}, corresponds to equal damping rates for cavity and qubits which are below
but comparable to the coupling strength $\gamma/\lambda = 0.2$. The entanglement between the two qubits
in the asymmetric detuning case can be enhanced when damping rate of the qubits are reduced.
Such a situation is shown on Fig.~\ref{SingletRWAvsNonRWA} for $\gamma / \lambda = 0.3$ and
$\gamma_s / \lambda = 3\times10^{-3}$. The total spin $\langle {\hat S}^2 \rangle$
then reduces to $\simeq 0.5$ at resonance indicating a higher degree of anti-locking
and compensation between the two qubits.

\begin{figure}[!htb]
\centering 
\includegraphics[width=0.6\columnwidth]{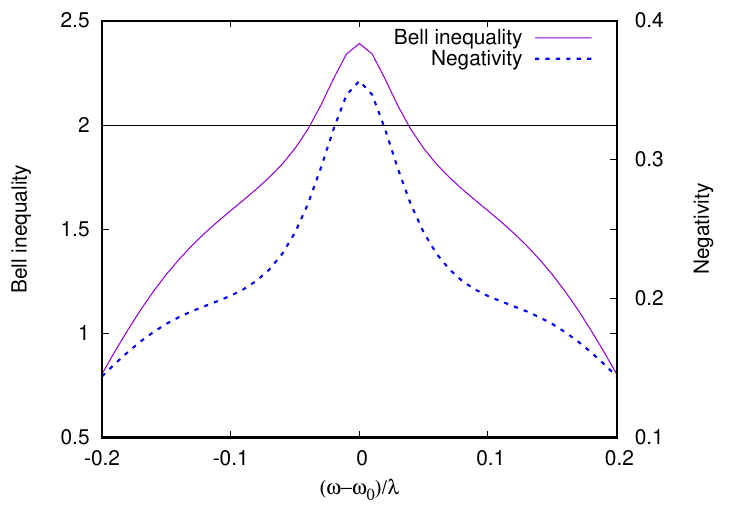} 
\caption{ Bell inequality violation and negativity of the steady-state RWA qubit pair with the reduced density matrix
  (trace done over the cavity) for the parameters of Fig.~\ref{fig8}. Since the qubit pair is in a mixture of
  singlet and triplet states, the polarization choice for the Bell inequality has to be adjusted to observe
  a Bell inequality violation (see  Appendix and also Fig.~\ref{figA2} there).
  Maximal negativity for two qubits is $1/2$ \cite{negativity} and thus
  this steady-state shows a high degree of stationary entanglement despite dissipative
  decoherence of both qubits and cavity.
}
\label{fig9}
\end{figure}

To confirm that this quantum synchronization and entanglement of the two qubits is not an artifact
of the RWA we performed direct simulations of the time dependent Lindblad dynamics of
Eqs.~(\ref{eq:lindblad}),~(\ref{eq:Ld}). Fig.~\ref{SingletRWAvsNonRWA}
presents a comparison between RWA and dynamical equations (\ref{eq:lindblad}),~(\ref{eq:Ld},
showing almost perfect agreement for $\omega/\Delta_1 = 30$.
Interestingly, even if larger deviations from RWA appear when  $\omega/\Delta_1 = 30$ is lowered,
the minimum $\langle {\hat S}^2 \rangle$ remains constant indicating that the anti-locking of
the qubits is robust to non-RWA effects which break the (anti)symmetry between the qubits shifting
the frequency at which minimum   $\langle {\hat S}^2 \rangle$ is achieved from exact resonance.
Even if the degree of cancellation between the qubits spins is not perfect it is sufficient
to generate steady-states violating Bell inequalities \cite{chuang}, this is shown in Fig.~\ref{fig9}.
In the calculation of the Bell inequality the spin projection has to be measured, in two directions by
two observers giving the freedom to choose four possible spin projection measurement directions for
the computation of the correlators. Since the entangled steady-state is not a pure singlet state we found
that violation of the Bell inequality was maximized with a $\sim 20^{\circ}$ rotation of the projection directions
compared to the rotation of $45^{\circ}$ which is used for Bell inequality tests on pure Bell (EPR) states.
To avoid any dependence on the projection angles, we also show the quantum negativity (see e.g. \cite{negativity}) of
the reduced qubit pair density matrix  (obtained from the total density matrix by tracing over the cavity).
The negativity at resonance in Fig.~\ref{fig9} is $0.36$, that is not far
from the maximal possible negativity of $0.5$ for a two qubit pair.
We emphasize that as opposed to entangled states generated by pulse sequences which have finite lifetimes,
this stationary entangled state is preserved in time in presence of dissipation and decoherence within our model.
The preservation of entangled state of two qubits
was also seen in numerical studies with quantum trajectories description
but there the dissipation was present only for the cavity and not for qubits \cite{zhirovqubit2}.

\begin{figure}[!htb]
\centering 
\begin{tabular}{cc}
\includegraphics[width=0.5\columnwidth]{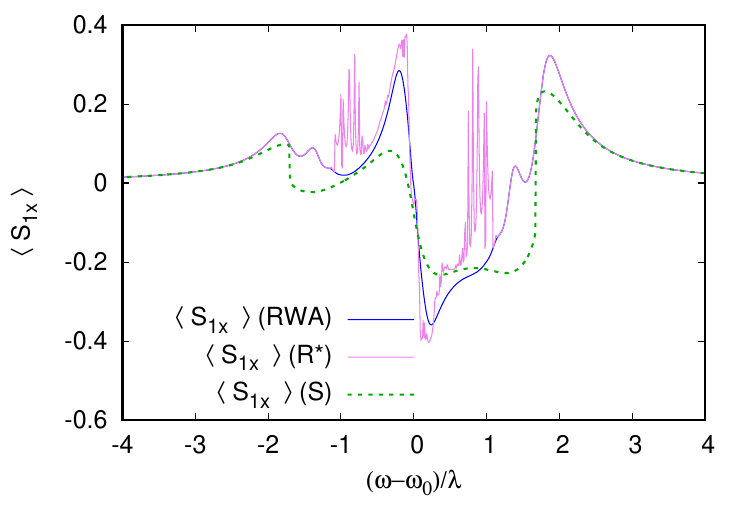} &
 \includegraphics[width=0.5\columnwidth]{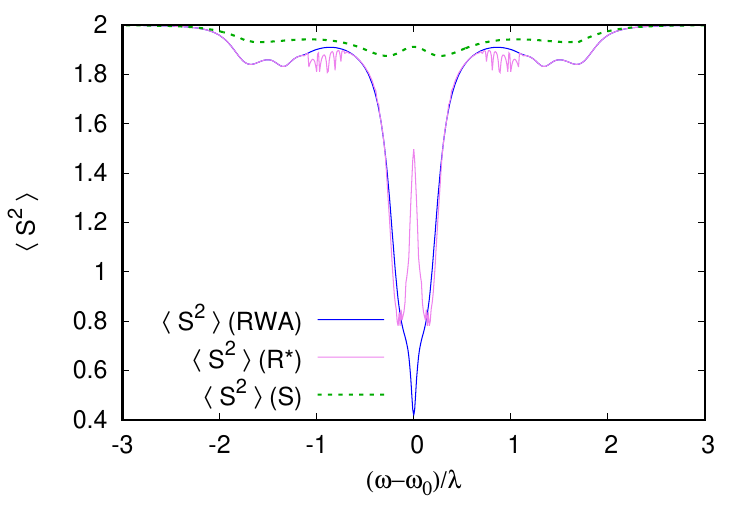} 
\end{tabular}
\caption{We test here if our two semi-analytic approaches can reproduce singlet formation
  for antisymmetric detuning presented in Figs.~\ref{fig8},\ref{fig9}. The left panel shows the RWA spin projection
  $\langle S_{1x} \rangle$ for the first qubit compared with the rate equation series and
  the variational approximation, the right panel shows the mean total spin $\langle S^2 \rangle$.
  While both approaches reproduce some qualitative features, they both fail
  to describe the singlet formation at $\omega-\omega_0 = 0$.
}
\label{FigSxBellTheo}
\label{fig10}
\end{figure}

To conclude this part we investigate if this anti-locked entangled state of the qubits could be reproduced
in the terms of the approximate rate equation or semiclassical method.
The rate equation series for $\Delta_1 = -\Delta_2$ fails to converge
in rather wide range $\omega_r/\Delta_1 \in (-1, 1)$ probably due to resonant energy levels
which are detrimental for convergence. Treating rate equations as an asymptotic series
and summing only the first terms which improve convergence allows to qualitatively reproduce
$\langle S_x \rangle$ in all the range $\omega_r$. However the prediction
for the total spin  $\langle {\hat S}^2 \rangle$ is then misleading
at resonance suggesting a maximum  $\langle {\hat S}^2 \rangle$ in contradiction with exact results,
as it is illustrated in Fig.~\ref{FigSxBellTheo}.
Not surprisingly the semiclassical approximation only succeeds
in reproducing qualitative features but completely misses the singlet formation. 

\section{Synchronization of several qubits}

\begin{figure}[!htb]
\centering 
\includegraphics[width=0.6\columnwidth]{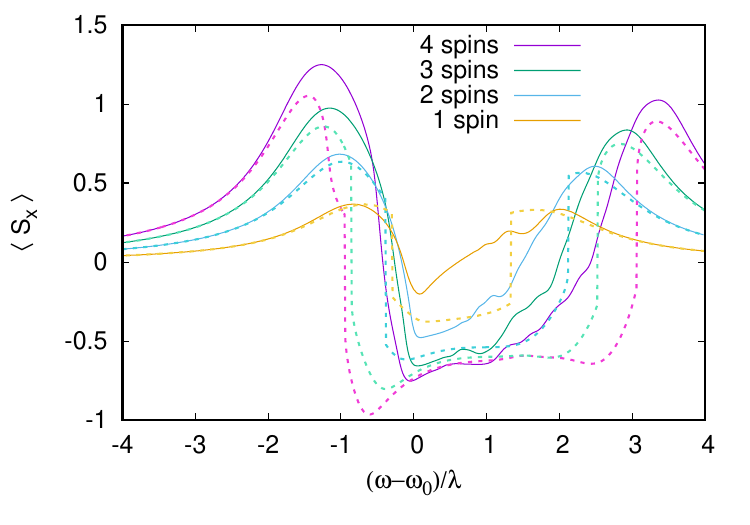}
\caption{RWA calculation of  total spin projection $\langle S_x \rangle$ of an increasing number of qubits
  coupled to one cavity (full curves) as function of cavity-excitation detuning and comparison
  to the semiclassical theory (dashed curves). The qubit cavity detunings are set to
  $\Delta_1 = \lambda$, $\Delta_2 = 1.5 \lambda$, $\Delta_3 = 2 \lambda$, $\Delta_4 = 2.5 \lambda$
  (only the first qubits are kept when the number of qubits is smaller than four). Dissipative rates are
  $\gamma = \gamma_S = 0.2 \lambda$ and excitation is $F = 0.77 \lambda$. Even if relaxation rates are all small,
  the semiclassical theory still captures many properties of $\langle S_x \rangle$, reproducing the increase
  of $\langle S_x \rangle$ with the number of qubits which corresponds to a synchronization of qubit rotation
  by the external drive. In both RWA and semiclassical data synchronization occurs
  at resonance $\omega = \omega_0$ but perhaps less expected at a higher frequency detuned from
  both cavity and qubits ($(\omega - \omega_0)/\lambda$ increasing from 2 to 3 with the number of qubits).
  Interestingly the accuracy of the semicalssical approximation seems
  to improve with the number of qubits, that may be due to its mean-field character.
}
\label{fig:severalqubits}
\label{fig11}
\end{figure}

\begin{figure}[!htb]
\centering 
\includegraphics[width=0.6\columnwidth]{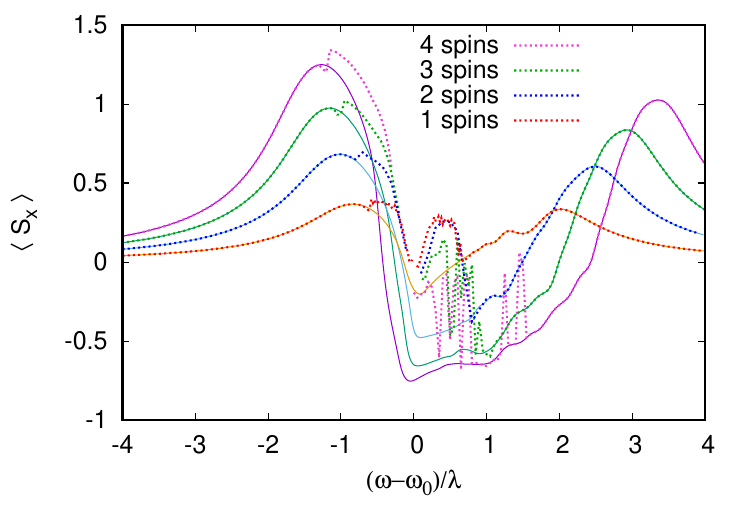}
\caption{The RWA calculation of  total spin projection $\langle S_x \rangle$ from Fig.~\ref{fig11}
  is compared to the result of the summation of the rate equation series (R)$^*$ (doted curves).
  The rate equation series give a result being almost exact away from resonance
  but they suffer from instabilities near
  $\omega = \omega_0$. It seems that the convergence range decreases with the number of
  qubits with various multi-qubit resonances making the series
  unstable even at $\omega - \omega_0 = 1.5 \lambda$ for four qubits. 
}
\label{fig:severalqubitsrate}
\label{fig12}
\end{figure}

It was shown in \cite{zhirovqubit1}, using the method of quantum trajectories, that the phase of rotating qubits
can be synchronized with the phase of microwave driving. We confirmed this by integration of
the quantum Lindblad evolution Eqs.~(\ref{eq:lindblad}),~(\ref{eq:Ld}) (see Appendix Fig.~\ref{figA3}).
We study this synchronization effect for up to four qubits. Our results are shown on Figs.~\ref{fig11},\ref{fig12} 
where the RWA steady-state is compared with semiclassical (Fig.~\ref{fig11}) and rate equation theories (Fig.~\ref{fig12}). 
All the qubits in this simulation are detuned from each other and without external driving their
in-plane magnetization will process at different frequencies leading to an average cancellation of
the total in plane magnetization. When the cavity is driven two synchronization peaks appear where
the total spin projection in the rotating frame $\langle S_x \rangle$ grows linearly with the number of detuned qubits.
The first peak is when the cavity is excited near resonance, the resonant cavity vibrations excite
the qubits which then all presses at the same phase and frequency. Surprisingly, a second synchronization peak appears
also at a higher frequency detuned from both cavity and qubit resonances.
The semiclassical approximation reproduces correctly the two synchronization peaks however the agreement
with RWA results is only qualitative. This is because this is a weak dissipation regime with strong cavity
qubit coupling $\Delta_1 = \lambda$ and $\gamma / \lambda = 0.2$. In Appendix Fig.~\ref{figA4}
we show that the agreement becomes almost perfect for the simpler weak coupling strong dissipation limit.
We note that in this simple regime the linear growth of $\langle S_x \rangle$ with the number of qubits
is observed only at cavity resonance. Fig.~\ref{fig11} suggests that the agreement between RWA and
semiclassical approximation tends to improve with the number of qubits, this maybe due to the fact
that our semiclassical approximation can also be viewed as a mean-field theory
whose accuracy improves with more interacting qubits.
The summation of the rate equation series reproduces the RWA result exactly away from cavity resonance
on Fig.~\ref{fig12}, however it seems that the diverging region around
resonance grows slowly with the number of qubits.

To summarize we find that this regime of quantum synchronization, where the total rotating in plane spin grows
with the number of qubits, is well described by both semiclassical and rate equation approaches.
As discussed in the previous Section, it is also possible to synchronize entangled qubits for antisymmetric qubit-cavity detunings in a vicinity of resonance between microwave driving and cavity frequency.

\section{Discussion}

In this work we used the Lindblad equation density matrix formalism
for a description of dissipative monochromatically driven resonator (oscillator cavity)
interacting with one or several qubits with very long or moderate
dissipative life time. This system extends the seminal Jaynes-Cummings model \cite{jc}
to a dissipative regime which is typical for superconducting qubits coupled
to a driven cavity \cite{wendin}. The performed numerical simulations
show that the results for the full time-dependent Lindblad equation (\ref{eq:lindblad}),~(\ref{eq:Ld})
are well reproduced by the solution of stationary RWA Lindblad equation (\ref{eq:RWAlindblad}).
Efficient exact  methods are developed to find the RWA steady-stae numerically with
several qubits and driven frequency being close to the cavity resonance
when many cavity eigenstates are excited. For one qubit case
a regime of qubit bistability is established  confirming previous results obtained with
the method of quantum trajectories \cite{zhirovqubit1}.

We also developed and tested two semi-analytical approaches
which allow to obtain the approximate accurate solutions
of exact RWA steady-state density matrix in the
regimes of relatively weak and strong dissipation
corresponding respectively to the rate equation and semiclassical approximations
for the steady-state density matrix. The analytical steps of these two approaches
significantly simplify the equations for density matrix
with consecutive numerical solution
and allow to investigate the system behavior
in a vicinity of cavity resonance with many
excited states. The numerical verification of these semi-analytical approaches
is done by the comparison with the exact RWA solution
and confirms their validity for a broad range of system parameters.
At the same time we demonstrate the existence of system behavior which
cannot be described by these semi-analytical approaches.
Thus we find that for two dissipative qubits and dissipative
driven cavity there exist a regime when qubits remain entangled,
forming a singlet, in the steady-state density matrix.
The existence of such a stationary regime of
entangled synchronized qubits, created by a monochromatically driven cavity, 
is really  surprising since both qubits and cavity are dissipative.
Due to the preservation of entanglement of qubits
such a regime can be considered as a real
entangled quantum synchronization.

We also find regimes (see Fig.~\ref{fig:severalqubits})
when up to 4 qubits are synchronized
with a phase of monochromatic cavity driving  
so that the total spin of the system grows proportionally
to the number of system qubits (spin halves).
At the same time we show that this synchronized
regime is well described by the semiclassical approach
so that one can discuss if such synchronized
regime of several qubits can be considered as purely
quantum or more as a regime of semiclassical synchronization
in presence of strong dissipation and noise induced by quantum fluctuations.
Indeed, it is known that the classical synchronization
is preserved in presence of moderate noise \cite{pikovsky}.
Even if this synchronization of several
qubits can be described in the frame of
semiclassical approach and that there is no entanglement
one can also argue that spin halves are purely quantum two-level
systems and hence their synchronization is also quantum.
The regime of entangled quantum synchronization
of qubits is purely quantum and cannot be obtained
in the frame of semiclassical description.

The obtained results provide a better understanding of
nontrivial behavior regimes of several
dissipative qubits interacting
with dissipative driven cavity. We hope that the developed methods
can be useful in other contexts.

\noindent {\bf Acknowledgments:}
The authors acknowledge support from the grants
 ANR France project OCTAVES (ANR-21-CE47-0007),
NANOX $N^\circ$ ANR-17-EURE-0009 in the framework of 
the Programme Investissements d'Avenir (project MTDINA),
MARS (ANR-20-CE92-0041) and EXHYP (INP Emergence 2022).

\clearpage

\section{Appendix}

\subsection{Semiclassical theory for a cavity coupled to several qubits}

We start by reminding the RWA Hamiltonian of a qubit coupled to two cavities:
\begin{align}
  {\hat H} &= \omega_r {\hat a}^+ {\hat a}  + F ( {\hat a} + {\hat a}^+ ) + \lambda_1 ( {\hat a} {\hat \sigma_1}^+ + {\hat a}^+ {\hat \sigma_1}^-) + \frac{\Omega_{r1}}{2} {\hat \sigma}_{1z}  + \lambda_2 ( {\hat a} {\hat \sigma_2}^+ + {\hat a}^+ {\hat \sigma_2}^-) + \frac{\Omega_{r2}}{2} {\hat \sigma}_{2z} \label{Hlambda2}
  \end{align}
where $\omega_r = \omega_0 - \omega$ and $\Omega_{r1} = \Omega_{1} - \omega, \Omega_{r2} = \Omega_{2} - \omega$.

We want to analytically compute and then minimize:
\begin{align}
S_2 = {\rm Tr}\;\left[{\cal L}({\hat \rho})\right]^+{\cal L}({\rho})
\end{align}
over trial density matrices which are given by the semiclassical ansatz:
\begin{align}
{\hat \rho} = \frac{1}{4} |\alpha\rangle \langle \alpha|\left(1 + \rho_{1x} {\hat \sigma}_{1x} + \rho_{1y} {\hat \sigma}_{1y} + \rho_z {\hat \sigma}_{1z} \right) \left(1 + \rho_{2x} {\hat \sigma}_{2x} + \rho_{2y} {\hat \sigma}_{2y} + \rho_z {\hat \sigma}_{2z} \right)
  \end{align}

For the Hamiltonian part of the functional we find:
\begin{align}
u_1 &= (1 + \rho_{1x}^2  + \rho_{1y}^2  + \rho_{1z}^2)/2 \label{u1def} \\
u_2 &= (1 + \rho_{2x}^2  + \rho_{2y}^2  + \rho_{2z}^2)/2 \label{u2def} \\
S_{0{\cal H}} &= 2(F^2 + 2 F \alpha_x \omega_r + (\alpha_x^2 + \alpha_y^2)\omega_r^2\\ \nonumber
S_{\lambda{\cal H}}(\rho_x,\rho_y,\rho_z,\Omega_r) &= 2 \lambda \left[ F \rho_x + (\alpha_x \rho_x - \alpha_y \rho_y) (\omega_r - \rho_z \Omega_r) \right] \\ 
& +\frac{\lambda^2}{2} \left[4 \rho_z^2 \left(\alpha_x^2+\alpha_y^2\right)+\left(4 \alpha_x^2+1\right) \rho_y^2+8 \alpha_x \alpha_y \rho_x \rho_y+\left(4 \alpha_y^2+1\right) \rho_x^2+(\rho_z+1)^2\right]  \\  \nonumber
S_{2{\cal H}} &= S_{0{\cal H}} u_1 u_2 + \frac{\Omega_{r1}^2(\rho_{1x}^2 + \rho_{1y}^2)}{2} u_2 + \frac{\Omega_{r2}^2(\rho_{2x}^2 + \rho_{2y}^2)}{2} u_1 \\
& + u_2 S_{\lambda_1{\cal H}}(\rho_{1x},\rho_{1y},\rho_{1z},\Omega_{r1}) + u_1 S_{\lambda_2{\cal H}}(\rho_{2x},\rho_{2y},\rho_{2z},\Omega_{r2}) + \lambda_1 \lambda_2 (\rho_{1x} \rho_{2x} + \rho_{1y} \rho_{2y})
\end{align}

To write the full functional we need to introduce also:
\begin{align} \\
S_0 &=  2 [F^2 + 2 F \alpha_x \omega_r + (\alpha_x^2 + \alpha_y^2) \omega_r^2] + 2 F \alpha_y \gamma + \gamma^2 \frac{\alpha_x^2 + \alpha_y^2}{2} \\
  S_{\gamma_{1S}} &= - \gamma_{1S} \lambda_1 (\alpha_y \rho_{1x} + \alpha_x \rho_{1y}) (2 + \rho_{1z})  + \gamma_{1S}^2 \frac{\rho_{1x}^2 + \rho_{1y}^2 + 4 (1 + \rho_{1z})^2}{8} \\
  S_{\gamma_{2S}} &= - \gamma_{2S} \lambda_2 (\alpha_y \rho_{2x} + \alpha_x \rho_{2y}) (2 + \rho_{2z})  + \gamma_{2S}^2 \frac{\rho_{2x}^2 + \rho_{2y}^2 + 4 (1 + \rho_{2z})^2}{8}   
\end{align}
The full semiclassical functional for a cavity coupled to two qubits is then given by:
\begin{align}
S_2 &= S_{0} u_1 u_2 + \frac{\Omega_{r1}^2(\rho_{1x}^2 + \rho_{1y}^2)}{2} u_2 + \frac{\Omega_{r2}^2(\rho_{2x}^2 + \rho_{2y}^2)}{2} u_1 \nonumber \\
& + u_2 S_{\lambda_1{\cal H}} + u_1 S_{\lambda_2{\cal H}} + \lambda_1 \lambda_2 (\rho_{1x} \rho_{2x} + \rho_{1y} \rho_{2y}) + \gamma u_2 \lambda_1 (\alpha_y \rho_{1x} + \alpha_x \rho_{1y}) +  \gamma u_1 \lambda_2 (\alpha_y \rho_{2x} + \alpha_x \rho_{2y}) \nonumber \\
&+ u_2 S_{\gamma_{1S}} +  u_1 S_{\gamma_{2S}} + \frac{\gamma_{1S} \gamma_{2S} \left[ \rho_{1x}^2 + \rho_{1y}^2 + 2 \rho_{z1} (1 + \rho_{1z})\right] \left[ \rho_{2x}^2 + \rho_{2y}^2 + 2 \rho_{z2} (1 + \rho_{2z}) \right] }{8} 
  \end{align}

The functional is then minimized over cavity parameters $\alpha_x, \alpha_y$ and spin projections $\rho_{1x}, \rho_{1y}, \rho_{1z}, \rho_{2x}, \rho_{2y}, \rho_{2z}$ using the NLopt optimization library. 

The generalization of this functional to more qubits is a direct generalization of the two-qubit functional keeping track all the terms which are generated by the possible qubit combinations, for example the term $S_{0} u_1 u_2$ becomes $S_{0} u_1 u_2 u_3 u_4$ for four qubits, $\lambda_1 \lambda_2 (\rho_{1x} \rho_{2x} + \rho_{1y} \rho_{2y})$ becomes $\lambda_1 \lambda_2 u_3 u_4 (\rho_{1x} \rho_{2x} + \rho_{1y} \rho_{2y})$ and so forth (the variables $u_{3,4}$ are defined in analogy with Eqs.~(\ref{u1def},\ref{u2def}). 

\subsection{Supplementary numerical results}

\begin{figure}[!htb]
  \centering
\begin{tabular}{cc}
\includegraphics[width=0.4\columnwidth]{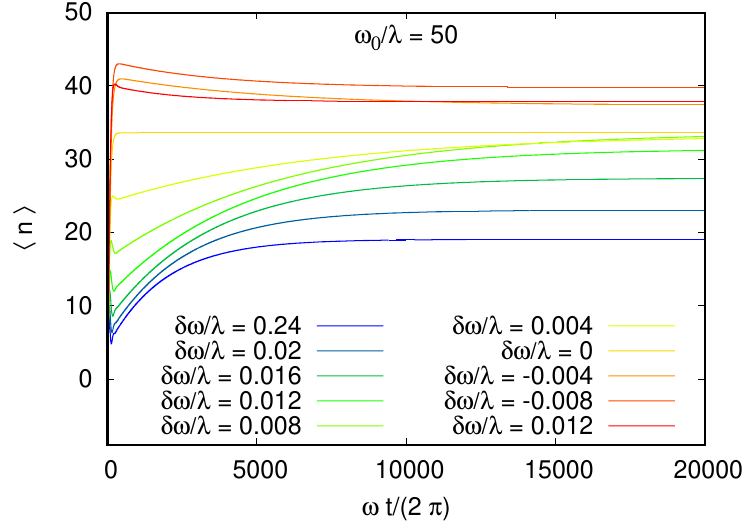} & \includegraphics[width=0.4\columnwidth]{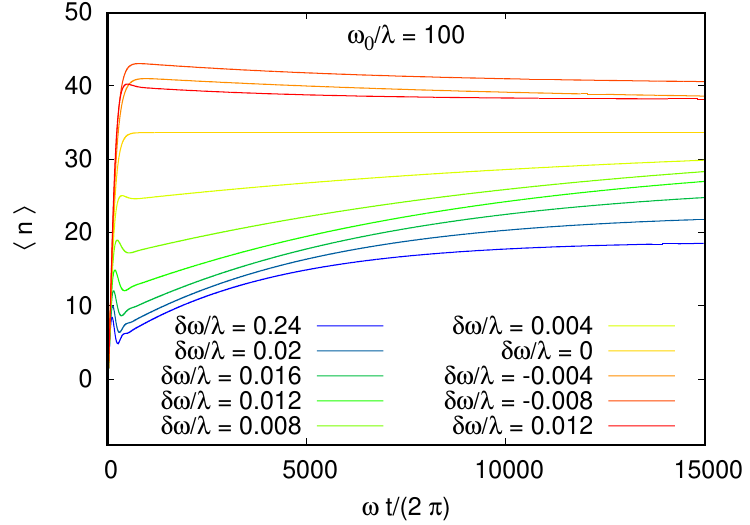}
\end{tabular}
\caption{Mean oscillator quantum number $\langle n \rangle = \langle a^+ a\rangle$ as function of
  the number of driving field oscillation periods $\omega t/(2 \pi)$ obtained by integration of
  the time dependent Lindblad equation for the data in Fig.~2 (we remind that
  $\Omega - \omega_0 = 2 \lambda$, $F = \lambda$, $\gamma = 0.03 \lambda$).
  The different curves correspond to different values of $\delta \omega = \omega - \omega_0$.
  The left panel shows $\omega_0/\lambda = 50$ and the right one  $100$. Relaxation is slower for
  positive values of $\delta \omega$ near resonance compared to negative $\delta \omega$.
  The comparison with RWA allowed us notice the incomplete relaxation in the simulations
  for $\omega_0/\lambda = 100$ which is easy to miss because it occurs in a narrow range of $\delta \omega$. 
}
\label{figA1}
\end{figure}

\begin{figure}[!htb]
\centering 
\includegraphics[width=0.6\columnwidth]{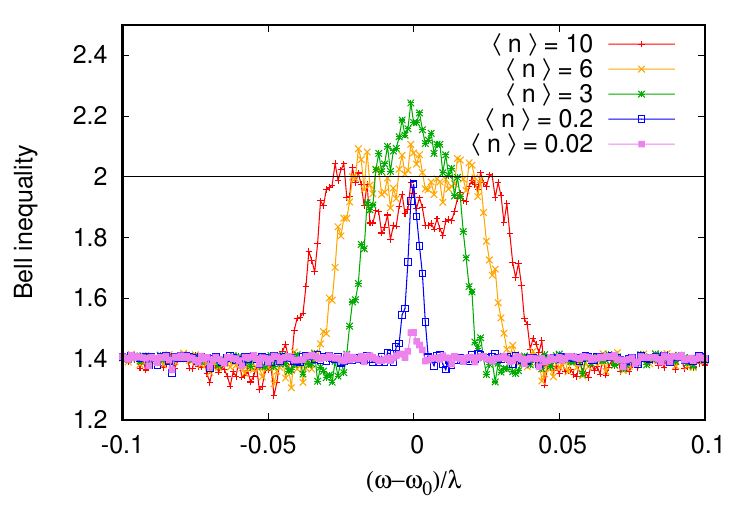}
\caption{ Violation of Bell inequality for the RWA steady-state for antisymmetric cavity qubit detunings
  $\Delta_1/\lambda = -\Delta_2 / \lambda = 0.5$ (see also Fig.~\ref{fig9}). Different curves correspond to
  increasing driving fields, expressed as the mean cavity occupation number at resonance
  $\langle n \rangle = 4 F^2/\gamma^2$ which is shown in the legend. Dissipative rates
  are $\gamma/\lambda = 0.025$ and $\gamma_s / \gamma = 0.02$. Maximal violation of Bell inequality
  is observed for $  \langle n \rangle = 3$ (at resonance). Further increase of the driving field
  leads to a reduction of Bell inequality violation highlighting the delicate
  quantum nature of the singlet state. To determine the violation of Bell inequalities,
  a maximum was taken over randomly chosen sets of 4 spin projection directions for the correlation measurement.
}
\label{figA2}
\end{figure}

\begin{figure}[!htb]
  \centering
\begin{tabular}{cc}
\includegraphics[width=0.4\columnwidth]{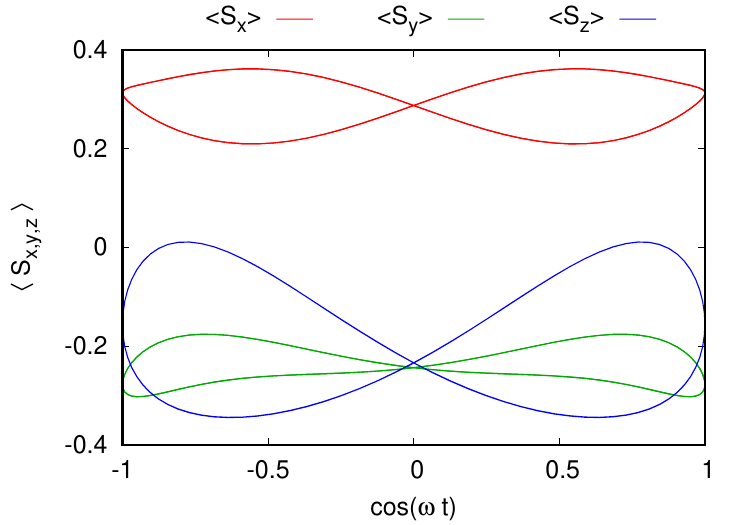} & \includegraphics[width=0.4\columnwidth]{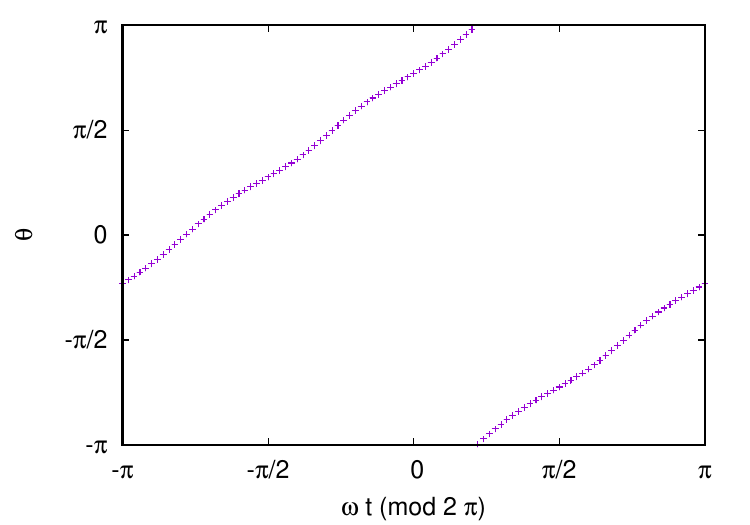}
\end{tabular}
\caption{Synchronization between a driven cavity and a qubit in the system steady-state for $\Delta_1 = \Omega_1 - \omega_0 = \lambda$,
  $F = \lambda$, $\gamma = \lambda/3$ and RWA parameter $\omega_0/\lambda = 10$. The qubit is non dissipative $\gamma_s = 0$
  and data is obtained by integration of Lindblad dynamics. Due to the moderate value of the RWA parameter vibrations around
  the mean RWA values of the spin projections are clearly visible on the left panel.
  Right hand panel shows the synchronlisation between the angle $\theta = \arg \langle S_x + i S_y \rangle$
  of the qubit in plane spin projection and the phase of the cavity driving field.  }
\label{figA3}
\end{figure}

\begin{figure}[!htb]
  \centering
\includegraphics[width=0.6\columnwidth]{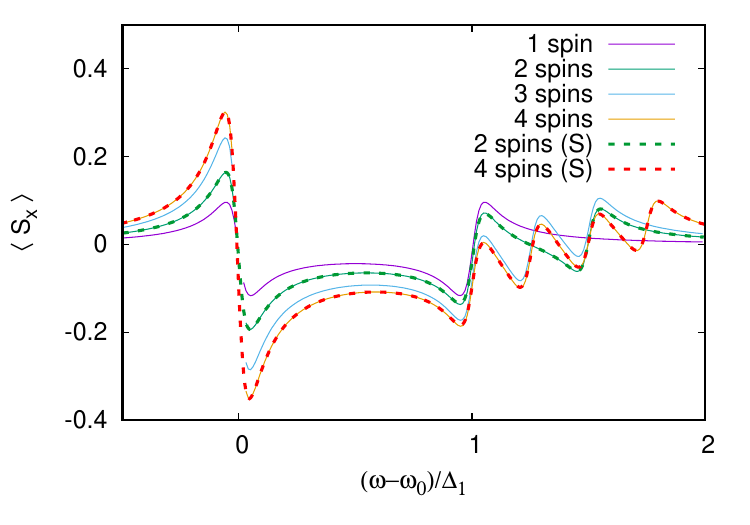}
\caption{RWA calculation of  total spin projection $\langle S_x \rangle$ as function of the excitation frequency-cavity
  detuning for an increasing number of qubits coupled to the cavity (full lines).
  The cavity-qubit interaction is weak $\lambda = \gamma/2$. The qubit-cavity detunings are set to $\Delta_1 = 10 \gamma^{-1}$,
  $\Delta_2 = 15 \gamma^{-1}$, $\Delta_3 = 12.5 \gamma^{-1}$, $\Delta_4 = 17.5 \gamma^{-1}$ (only the first qubits
  are kept when the number of qubits is smaller than four). Qubit dissipation rate is $\gamma_S = \gamma$ and excitation
  is $F/\gamma = 2.2$. In this weak interaction regime the semiclassical calculation
  (shown as dashed lines for 2 and 4 spins) coincides almost exactly with RWA. }
\label{figA4}
\end{figure}

Here we provide some additional numerical data supporting the findings presented in the main text. In Fig.~\ref{figA1} we show the slow relaxation for some values of the detuning $\omega - \omega_0$ in the model where the qubit is not dissipative $\gamma_s = 0$. In Fig.~\ref{figA2} we investigate how the maximal violation of Bell inequalities for antisymmetric qubit-cavity detuning depends on cavity driving strength.  Fig.~\ref{figA3} shows synchronized cavity-qubit behavior and oscillations around the RWA steady state due to finite values of the RWA parameter $\omega_0/\lambda$. Fig.~\ref{figA4} compares the RWA steady state with the semiclassical theory for several qubits in the regime of weak qubit-cavity interaction.

\clearpage



\begin{thebibliography}{99}
\bibitem{huygens} C.~Huygens, {\it {\OE}vres compl\'etes},
  vol. 15, Swets \& Zeitlinger B.V., Amsterdam (1967).
\bibitem{pikovsky} A.~Pikovsky, M.~Rosenblum and J.~Kurths,
     {\it Synchronization: a universal concept in nonlinear sciences},
     Cambridge University Press, Cambridge UK (2001).
\bibitem{weisenfeld} M.~Bennett, M.F.~Schatz, H.~Rockwood and 
     K.~Wiesenfeld, {\it Huygen's clocks}, Proc. R. Soc. Lond. A {\bf 458}, 563 (2002).
\bibitem{likharev} A.K.~Jain, K.K.~Likharev, J.E.~Lukens
     and J.E.~Sauvageau, {\it Mutual phase-locking in Josephson junction arrays},
     Phys. Rep. {\bf 109}, 309 (1984).
\bibitem{chuang} M.A.~Nielsen and I.L.~Chuang,
     {\it Quantum computation and quantum information},
            Cambridge Univ. Press, Cambridge UK (2000).
\bibitem{wendin} G.~Wendin, {\it Quantum information processing 
        with superconducting circuits: a review},
        Rep. Prog. Phys. {\bf 80}, 106001 (2017).
\bibitem{wallraff1} A.~Wallraff, D.~I.~Schuster, A.~Blais, L.~Frunzio, 
     R.~S.~Huang, J.~Majer, S.~Kumar, S.~M.~Girvin and R.~J.~Schoelkopf,
      Nature {\bf 431}, 162 (2004).
\bibitem{wallraff2} J.~Majer, J.~M.~Chow, J.~M.~Gambetta, J.~Koch, B.~R.~Johnson, 
   J.~A.~Schreier, L.~Frunzio, D.~I.~Schuster, A.~A.~Houck, A.~Wallraff, 
   A.~Blais, M.~H.~Devoret, S.~M.~Girvin and R.~J.~Schoelkopf,
   {\it Strong coupling of a single photon to a superconducting qubit using circuit quantum electrodynamics},
    Nature {\bf 449}, 443 (2007).
\bibitem{wallraff3} J.~M.~Fink, M.~G\"oppl, M.~Baur, R.~Bianchetti, P.~J.~Leek, 
  A.~Blais and A.~Wallraff,
  {\it Climbing the Jaynes–Cummings ladder and observing its nonlinearity in a cavity QED system },
  Nature {\bf 454}, 315 (2008).
\bibitem{wallraff4} S.~Filipp, P.~Maurer, P.~J.~Leek, M.~Baur, R.~Bianchetti, 
  J.~M.~Fink, M.~G\"oppl, L.~Steffen, J.~M.~Gambetta, A.~Blais and A. Wallraff,
  {\it Two-qubit state tomography using a joint dispersive readout},
   Phys. Rev. Lett. {\bf 102}, 200402 (2009).
\bibitem{wallraff5} J.~M.~Fink, R.~Bianchetti, M.~Baur, M.~Goeppl, L.~Steffen, 
   S.~Filipp, P.~J.~Leek, A.~Blais and A.~Wallraff,
   {\it Dressed Collective Qubit States and the Tavis-Cummings Model in Circuit QED},
   Phys. Rev. Lett. {\bf 103}, 083601 (2009).
\bibitem{gorini} V.~Gorini, A.~Kossakowski and E.C.G.~Sudarshan,
  {\it Completely positive dynamical semigroups of N-level systems},
  J.Math.Phys. {\bf 17}, 821 (1976).
\bibitem{lindblad} G.~Lindblad,
  {\it On the generators of quantum dynamical semigroups},
  Commun. Math. Phys. {\bf 48},119 (1976).
\bibitem{weiss} U.~Weiss, {\it Quantum dissipative systems}, 5th ed.,
  World Scieentific, Singapore (2021).
\bibitem{graham1987}  T.~Dittrich and R.Graham,
  {\it Quantum effects in the steady state of the dissipative standard map},
  Europhys. Lett. {\bf 4(3)}, 263 (1987).
\bibitem{graham1988}  T.~Dittrich and R.Graham,
  {\it Effects of weak dissipation on the long-time behaviour of the quantized standard map},
   Europhys. Lett. {\bf 7(4)}, 287 (1988).
\bibitem{brun1996} T.A.~Brun, I.C.~Percival and R.~Schack,
  {\it  Quantum chaos in open systems: a quantum state diffusion analysis},
  J. Phys. A: Math. Gen. {\bf 29}, 2077 (1996).
\bibitem{brun2002}  T.A.~Brun, {\it A simple model of quantum trajectories},
  Am. J. Phys. {\bf 70}, 719 (2002).
\bibitem{zhirovqsync} O.V.~Zhirov and D.L.~Shepelyansky,
  {\it Quantum synchronization},
  Eur. Phys. J. D {\bf 38}, 375 (2006).
\bibitem{zhirovqubit1} O.V.~Zhirov and D.L.~Shepelyansky,
  {\it Synchronization and bistability of qubit coupled to a driven dissipative oscillator},
  Phys. Rev. Lett. {\bf 100}, 014101 (2008).
\bibitem{zhirovqubit2} O.V.~Zhirov and D.L.~Shepelyansky,
  {\it Quantum synchronization and entanglement of two qubits coupled to a driven dissipative resonator},
  Phys. Rev. B {\bf 80}, 014519 (2009). 
\bibitem{pikovdrive} A.S.~Pikovsky, M.G.~Rosenblum, G.V.~Osipov and J.~Kurths,
  {\it Phase synchronization of chaotic oscillators by external driving},
  Physica D {\bf 104}, 219 (1997).
\bibitem{fazio} A. Mari, A. Farace, N. Didier, V. Giovannetti and R. Fazio,
  {\it Measures of Quantum Synchronization in Continuous Variable Systems},
  Phys. Rev. Lett. {\bf  111}, 103605 (2013).
\bibitem{bruder1} S.~Walter, A.~Nunnenkamp, and C.~Bruder,
               {\it Quantum synchronization of a driven self-sustained oscillator},
               Phys. Rev. Lett. {\bf 112}, 094102 (2014).
\bibitem{holland} M.~Xu, D.A.~Tieri, E.C.~Fine, J.K.~Thompson and M.J.~Holland,
  {\it Synchronization of two ensembles of atoms},
  Phys. Rev. Lett. {\bf 113}, 154101 (2014).
\bibitem{mavrogordatos} Th.K.~Mavrogordatos, G.~Tancredi, M.~Elliott, 
             M.J.~Peterer, A.~Patterson, J.~Rahamim,  P.J.~Leek, 
             E.~Ginossar, and M.H.~Szymanska,
             {\it Simultaneous bistability of a qubit and 
              resonator in circuit quantum electrodynamics},
             Phys. Rev. Lett. {\bf 118}, 040402 (2017).
\bibitem{nori} X.~Guab, A.F.~Kockum, A.~Miranowicz, Y-x.~Liu,
          and F.~Nori,
         {\it Microwave photonics with superconducting quantum circuits},
          Phys. Rep. {\bf 718-719}, 1 (2017).
\bibitem{bruder2} A.~Roulet, and C.~Bruder,
                {\it Synchronizing the smallest possible system},
                   Phys. Rev. Lett. {\bf 121}, 053601 (2018).
\bibitem{bruder3} A.~Roulet, and C.~Bruder,
         {\it Quantum synchronization and entanglement generation},
         Phys. Rev. Lett. {\bf 121}, 063601 (2018).
\bibitem{satellite} S.~Swaraj, O.~Lhamo,  M.~Paul, R.~Bassoli and F.H.P.~Fitzek,
         {\it Quantum time synchronization for satellite networks},
         doi.org/10.36227/techrxiv.22325068.v1 (2023).
\bibitem{satnat2023} E.D.~Caldwell, J.-D.~Deschenes, J.~Ellis, W.C.~Swann, B.K.~Stuhl,
        H.~Bergeron, N.R.~Newbury and L.C.~Sinclai,
        {\it Quantum-limited optical time transfer for future geosynchronous links},
        Nature {\bf 618}, 721 (2023).
\bibitem{satnat2023b}   D.~Gozzard,
           {\it Clocks synchronized at the quantum limit},
         Nature {\bf 618}, 680 (2023).
\bibitem{jc} E.T.~Jaynes, and F.W.~Cummings,
         {\it Comparison of quantum and semiclassical radiation theories 
           with application to the beam maser"},  Proc. IEEE. {\bf 51(1)}, 89 (1963).
\bibitem{eberlybook} L.~Allen and J.H.~Eberly,
        {\it Optical resonance and two-level atoms},
            Dover Publs. Inc., New York (1987).
\bibitem{scully} M.O.~Scully, and M.S.~Zubairy, 
          {\it Quantum optics},
            (Cambridge University Press, Cambridge, England, 1997).
\bibitem{walther} G.~Rempe, H.~Walther and N.~Klein, 
         {\it Observation of quantum collapse and revival in a one-atom maser}, 
         Phys. Rev. Lett. {\bf 58(4)}, 353 (1987).
\bibitem{jcdriven} L.~Ermann, G.G.~Carlo, A.D.~Chepelianskii and D.L.~Shepelyansky,
          {\it Jaynes-Cummings model under monochromatic driving},
           Phys. Rev. A {\bf 102}, 033729 (2020).
\bibitem{sylvester} R.H.~Bartels and G.W.~Stewart,
         {\it Solution of the matrix equation AX+XB=C},
         Comm. ACM {\bf 15(9)}, 820 (1972).
\bibitem{King-wah} Eric King-wah, {\it The solution of the matrix equations AXB-CXD= E AND (YA-DZ, YC-BZ)=(E, F)},
         Linear Algebra and its Applications, {\bf 93}, 93 (1987)
\bibitem{odeint} ODEINT Library,
        \url{https://headmyshoulder.github.io/odeint-v2/}
        (Accessed June 2023).
\bibitem{jacobi} Ke Chen, {\it  Matrix preconditioning techniques and applications},
          Cambridge Univ. Press, Cambridge UK (2005).        
\bibitem{negativity}  G.~Vidal and R.F.~Werner, {\it A computable measure of entanglement},
        Ohys. Rev. A {\bf 65},  032314 (2002).
        
\end{thebibliography}
\end{document}